\documentclass[acmsmall]{acmart}

\usepackage{csquotes}
\usepackage[export]{adjustbox}
\usepackage{caption}
\usepackage{subcaption}
\usepackage{tabularx}

\def\BibTeX{{\rm B\kern-.05em{\sc i\kern-.025em b}\kern-.08emT\kern-.1667em\lower.7ex\hbox{E}\kern-.125emX}}
    
\setcopyright{acmlicensed}
\acmJournal{PACMHCI}
\acmYear{2020} \acmVolume{4} \acmNumber{CSCW2} \acmArticle{168} \acmMonth{10} \acmPrice{15.00}\acmDOI{10.1145/3415239}

\begin{document}

%
\title{Good for the Many or Best for the Few? A Dilemma in the Design of Algorithmic Advice}

%
\author{Graham Dove}
\affiliation{%
  \institution{Tandon School of Engineering, New York University}
  \city{New York}
  \country{USA}}
\email{gd64@nyu.edu}

\author{Martina Balestra}
\affiliation{%
  \institution{Tandon School of Engineering, New York University}
  \city{New York}
  \country{USA}}

\author{Devin Mann}
\affiliation{%
 \institution{Grossman School of Medicine, New York University}
 \city{New York}
 \country{USA}}
 
\author{Oded Nov}
\affiliation{%
  \institution{Tandon School of Engineering, New York University}
  \city{New York}
  \country{USA}}
\email{onov@nyu.edu}

%
\renewcommand{\shortauthors}{Dove, et al.}

%
\begin{abstract}
Applications in a range of domains, including route planning and well-being, offer advice based on the social information available in prior users' aggregated activity. When designing these applications, is it better to offer: a) advice that if strictly adhered to is more likely to result in an individual successfully achieving their goal, even if fewer users will choose to adopt it? or b) advice that is likely to be adopted by a larger number of users, but which is sub-optimal with regard to any particular individual achieving their goal?  We identify this dilemma, characterized as \textit{Goal-Directed} vs. \textit{Adoption-Directed} advice, and investigate the design questions it raises through an online experiment undertaken in four advice domains (financial investment, making healthier lifestyle choices, route planning, training for a 5k run), with three user types, and across two levels of uncertainty. We report findings that suggest a preference for advice favoring individual goal attainment over higher user adoption rates, albeit with significant variation across advice domains; and discuss their design implications. 
\end{abstract}

%
%


%

%

%
\maketitle

\section{Introduction}
Applications that offer advice to users have become commonplace across domains as diverse as route planning (e.g. Google Maps or Citymapper), financial investment (e.g. PlanMode), and lifestyle and well-being (e.g. Noom). Increasingly, this advice is generated algorithmically, based on machine learning analysis of large amounts of historic data relating to aggregated prior use. The advice on offer thereby combines social cues from other users' activity with up-to-date data from the domain in question. Two important questions that may be considered when formulating this advice are: 1) is the user likely to adopt the advice, if it is offered? and 2) if adopted, is the advice likely to lead to the user achieving their goal? 

In many cases, offering the advice most likely to result in users achieving their goal will also result the highest rate of user adoption. However, this is not always the case. There are times when this advice may seem counter-intuitive; such as when traffic conditions result in a route much longer in distance being shorter in duration. In other situations, such as advising major changes in lifestyle associated with improving personal health, the `best' advice might appear exceptionally challenging.  For example, behavior in response to social distancing guidelines during the COVID-19 pandemic has highlighted the tension between the standalone value of public advice on one hand and its adoption by individuals who find it hard to follow on the other. Extant research shows that in cases like these people do not always act like models of rational decision making, and may not necessarily make decisions that maximize utility, or follow advice that is optimal to achieving their goal. Instead, their decisions are likely to be based on heuristics and judgments that follow predictable biases, e.g. \cite{tversky1974judgment, kahneman979prospect, kahneman2006anomalies, dellavigna2009psychology, thaler2018cashews, barberis2013thirty}. These circumstances pose a potentially serious design dilemma. Should the user be offered the advice which, if strictly adhered to, would be most likely to result in them achieving their objective, even if they are statistically less likely to adopt or adhere to it? Or alternatively, should the user be offered advice that they are more likely to choose to adopt and see through, and which may still be helpful, even though they are less likely to achieve their goal in full? We characterize this dilemma as being between: 1) \textit{Goal-Directed advice} that is more likely to lead a user who fully adopts it to achieve their goal, even if overall adoption rates are likely to be lower, and 2)  \textit{Adoption-Directed advice} which is likely to have a higher adoption rate, but which has a lower probability of resulting in a user fully achieving their individual goal.

This dilemma is brought even more clearly into focus when machine learning and data analysis techniques used to predict likely outcomes might also be used to predict user adoption rates. What might this mean for designers of applications aiming to algorithmically generate valuable advice in complex situations? Which design choice would be preferable from the perspective of a user? Which type of advice might be considered more ethical? This paper reports on a study in which participants in an online experiment were asked to indicate a preference between examples of these different advice types in one of four different scenarios, each presenting the potential dilemma in a different domain context with particular characteristics that influence how the dilemma might present. We varied the role participants were asked to adopt when indicating their preference, and controlled for several other potentially confounding factors. Our analysis found that participants consistently favored presenting Goal-Directed advice over Adoption-Directed advice, albeit the degree to which this preference was shown differed significantly across the scenarios that were presented. We discuss our findings with reference to prior research in decision-making under uncertainty, recommender systems, and behavioral economics, and outline implications for designing systems that offer algorithmically generated advice.

\subsection{The Advice Design Dilemma}

The research reported in this paper addresses a dilemma in presenting users with algorithmic advice that aims to support their decision making under uncertainty, particularly in challenging contexts. This dilemma can appear when advice, which may be optimal to an individual achieving their goal, appears counter-intuitive. Or when that advice seems challenging to such a degree that many users may either discount or ignore it. In these circumstances, is this advice still optimal? and should an application offer this advice? Or is it preferable for an application to offer advice that a higher percentage of users are more likely to adopt, but which is less likely to lead these users towards fully achieving their goal? 

For the purposes of this study we define these two alternative advice types as follows:

\begin{itemize}
    \item \textit{Goal-Directed advice:} advice that is more likely to result in users who adhere to it achieving their goal, but which users are less likely to adopt and adhere to.
    \item \textit{Adoption-Directed advice} advice that is more likely to be adopted and adhered to by a greater number of users, but which is less likely to result in a user fully achieving their goal.
\end{itemize}

It should be noted that in the situations discussed in this paper, we make explicit the assumption that selecting either of the advice options is preferable to acting independently, and would therefore result in a better outcome for the user than were they to follow neither.

\subsection{Contributions}
This research contributes to the CSCW community's understanding of how to design applications that offer algorithmically generated advice based on social information aggregated from the prior behavior of a large number of users, and has both theoretical and practical implications. From a theoretical perspective, we 1) extend prior research at the intersection of CSCW and recommender systems that offer socially-informed advice based on aggregated data from prior use;  2) introduce and characterize the design dilemma posed when having to select between Goal-Directed and Adoption-Directed advice; and 3) quantify the trade-offs it represents. From a practical perspective we offer initial guidance on how to approach this dilemma when designing applications that offer algorithmically generated advice.

\section{Related Work}
This research was inspired by the observation that users may discount or ignore algorithmic advice intended to support the choices they make towards achieving particular goals, specifically when this advice may appear counter-intuitive or challenging to adhere to. In order to situate our work we provide an overview of related prior research in 1) decision making under uncertainty; and 2) online advice and recommendation.

\subsection{Decision-making under uncertainty}
A primary motivation for offering algorithmic advice is to support users' decision-making. Under uncertainty, decision-makers tend to be receptive to advice \cite{bonaccio2006advice,yaniv2000advice}, from experts \cite{cialdini1993influence,jowett2018propaganda,milgram1963behavioral,tyler1992relational} and from peers \cite{eysenbach2004health,bursztyn2014understanding,thaler2007behavioral}. This tendency has informed the design of interfaces for recommender systems. For example, in financial planning  \cite{adomavicius2005toward,wang2007recommendation} advice from software agents is increasingly replacing advice from humans \cite{adomavicius2011context,jannach2016recommender}, evidenced in the growth of financial `Robo-Advisors', such as Betterment \footnote{www.betterment.com}, which provide online advice with minimal human intervention.

Even where valuable advice may be available, decision-makers still tend to rely on simple heuristic principles to reduce the complexity of tasks such as assessing probabilities and predicting values \cite{kahneman979prospect,barberis2013thirty}. While often useful, heuristics can also lead to systematic errors or biases \cite{tversky1974judgment}. Their impact has been extensively studied in the context of personal finance \cite{benartzi1999risk,thaler2007behavioral}. Prior research shows the effectiveness of changing default participation to opt-in, but also the negative impact a reliance on simplistic heuristics has on asset allocation \cite{benartzi2007heuristics}. Decision-makers may excessively discount advice even when shown that the advice is good \cite{goodwin1999judgmental,lim1995judgemental}, because they have access to their own but not others' justifications \cite{yaniv2000advice}, or because of a bias toward egocentric or self-related information \cite{kruger1999lake,krueger2001role,clement2000primacy,dunning1996evidence}. As a result, decision-makers tend to exaggerate their own abilities \cite{kruger1999lake}, not take others' skills sufficiently into account, and display overconfidence and unrealistic self-assessments \cite{block1991overconfidence}. However, task complexity and quality of explanation may provide a counter-balance that reduces advice discounting \cite{onkal2009relative}, and decision-making can be affected by context driven challenges to self-control, which may represent inconsistent long-term and short-term preferences \cite{shefrin1988behavioral,thaler2018cashews}.

\subsection{Online Advice and Recommender Systems}
Insights such as these, from behavioral economics and social psychology, have influenced HCI, CSCW and information systems research (e.g. \cite{lee2011mining,gunaratne2015informing,caraban201923}) in areas such as recommender systems \cite{swearingen2001beyond, adomavicius2019reducing}. Recommender systems present advice to users in the form of suggestions \cite{nguyen2015perverse, wang2018tem} often based on collaborative filtering and a social context  \cite{herlocker2000explaining, tsai2017providing}. Recommender systems have been used to provide advice in a wide range of domains and use cases \cite{jannach2016recommender}, including advice on energy saving \cite{azaria2014advice}, nutrition \cite{zenun2017online}, finance   \cite{gunaratne2018persuasive}, and remedial behavior for computer programmers \cite{hartmann2010would}. Users' interactions with such systems are susceptible to similar confirmation biases as other instances of decision-making and advice selection \cite{jonas2001confirmation}, which has led researchers to investigate diverse recommendation and dissenting information strategies \cite{buder2012learning,nov2013personality}. In addition, prior research at the intersection of CSCW and recommendation systems found that negative social effects of activity transparency can result in users' increased adoption of mediocre advice \cite{nguyen2015perverse}. Another significant challenge to overcome, identified in this line of research, is the tendency for users to be put off by recommendations they consider demanding or challenging \cite{schwind2011will}. Moreover, the recommendations users find most useful may not always be the ones that by other measures are considered most accurate \cite{mcnee2006being}. An example is when optimal advice appears counter-intuitive, but following sub-optimal advice will still result in increased profit or satisfaction \cite{levy2016intelligent}. We extend this prior research by investigating participants' preferences between Goal-Directed and Adoption-Directed advice, under uncertainty and in counter-intuitive or challenging scenarios.

\section{Experimental Study}
To investigate the advice design dilemma, we probe participants' design preferences for which advice type to present in a future app. To study this, we conducted a between-subjects experiment online using Amazon Mechanical Turk (MTurk) as a recruitment platform. While acknowledging that the dilemma may appear in situations where there are more complex, ongoing interactions with applications that reoccur over time, and involve both short-term and evolving longer-term goals, e.g. when using diet and exercise apps,  we chose to echo a long tradition in economics, where simple, discrete-choice experiments are used to isolate a particular issue of concern and elicit preferences,  e.g. \cite{brynjolfsson2019using,bansback2012using,egesdal2015estimating,horne2005improving}. This degree of simplification allows us to gain purchase on the novel conceptual ground that the dilemma reflects, by allowing us to isolate the particular cases where advice might appear counter-intuitive or challenging, and to investigate the impact of factors, such as task domain, on these instances.

\subsection{Research Questions}
We ask the following research questions:
\begin{itemize}
    \item \textit{RQ1: Are participants' preferences between offering Goal-Directed advice and Adoption-Directed advice sensitive to different domain scenarios, or do they transcend specific settings?}
    \item\textit{RQ2: Are participants' preferences between offering Goal-Directed advice and Adoption-Directed advice sensitive to the different perspectives of `advice giver' and `advice receiver'?}
\end{itemize}

\subsection{Study Design}
We recruited a total of 1,589 US MTurk participants over the age of 18. Participation was limited to people with a record of at least 100 tasks at an approval rate of 95\% or higher.  
Of these, 750 self-identified as women and 829 as men. Their self-reported ages ranged between 18 and 88, with a mean age of 36 (median = 33). Participants were paid a flat rate of \$0.65 for completing the survey based on an estimated time for completion of 5 minutes (median time spent on the study by participants was 5 minutes 30 seconds), and could take part in the study only once. Each participant was presented with a single selection task, asking for their preference between Goal-Directed and Adoption-Directed advice. Based on a 4X3 factorial design, participants were randomized into one of four scenarios, each set in a different domain, and three roles through which to consider their selection. We also controlled for other potentially confounding factors, such as the size of the gap between adoption and goal-effectiveness probabilities, and order of presentation. In addition to making their advice type selection, participants were required to answer an attention test question. Finally, participants were asked to briefly explain the reasons behind their selection. In the following sections we report the details of experimental procedure.

\subsection{Domain Scenarios}
\begin{figure}[t]
 \centering
  \begin{subfigure}{0.49\textwidth}
  \captionsetup{margin=0.1cm}
  \includegraphics[width=0.95\linewidth]{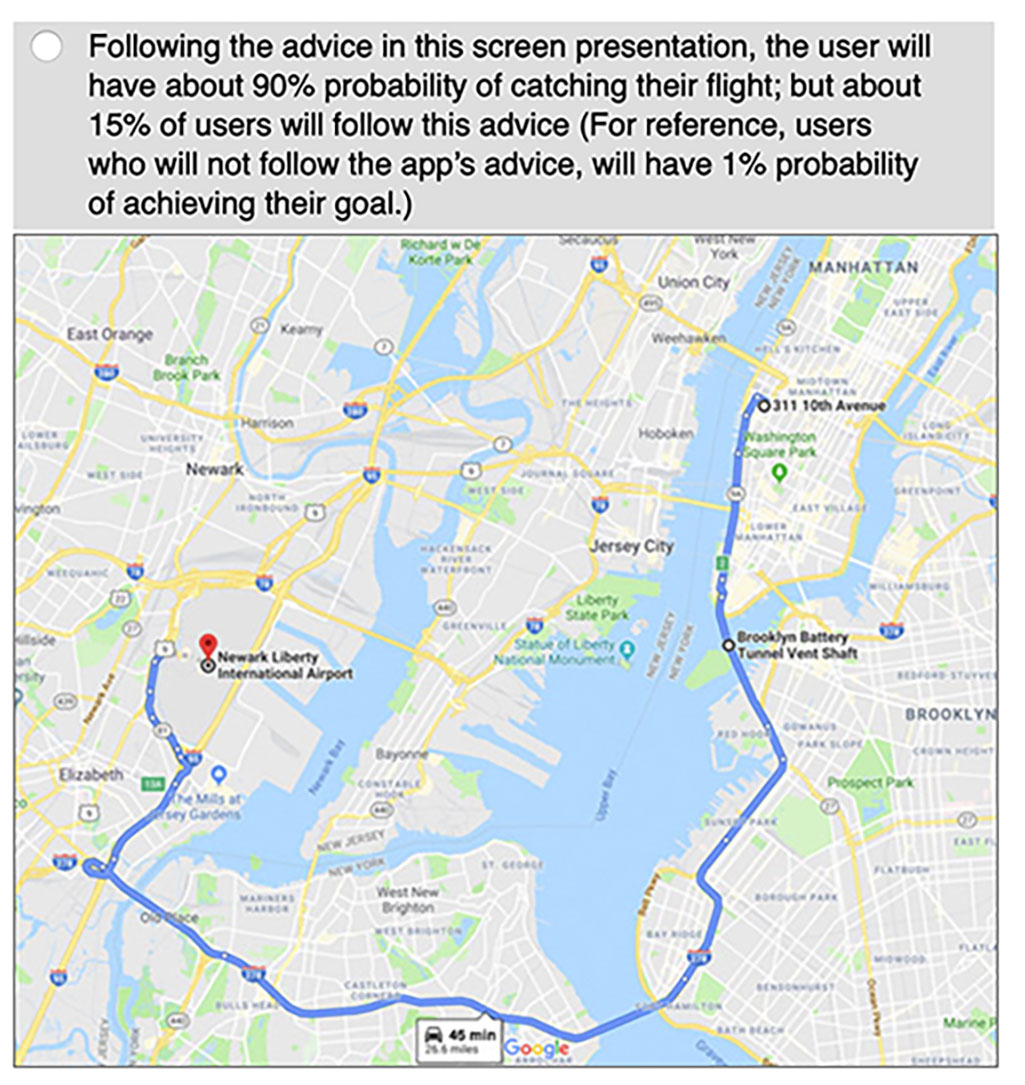} 
  \caption{`Goal-Directed' advice: Participants are shown a route that is longer in distance but shorter in duration. They are instructed that users will have a 90\% probability of catching their flight, but that only 15\% of users are likely to adopt it.}
  \label{fig:scenario1}
  \Description{Screenshot of the `Route planning' scenario, showing `Goal-directed' advice}
  \end{subfigure}
  \begin{subfigure}{0.49\textwidth}
  \captionsetup{margin=0.1cm}
  \includegraphics[width=0.95\linewidth]{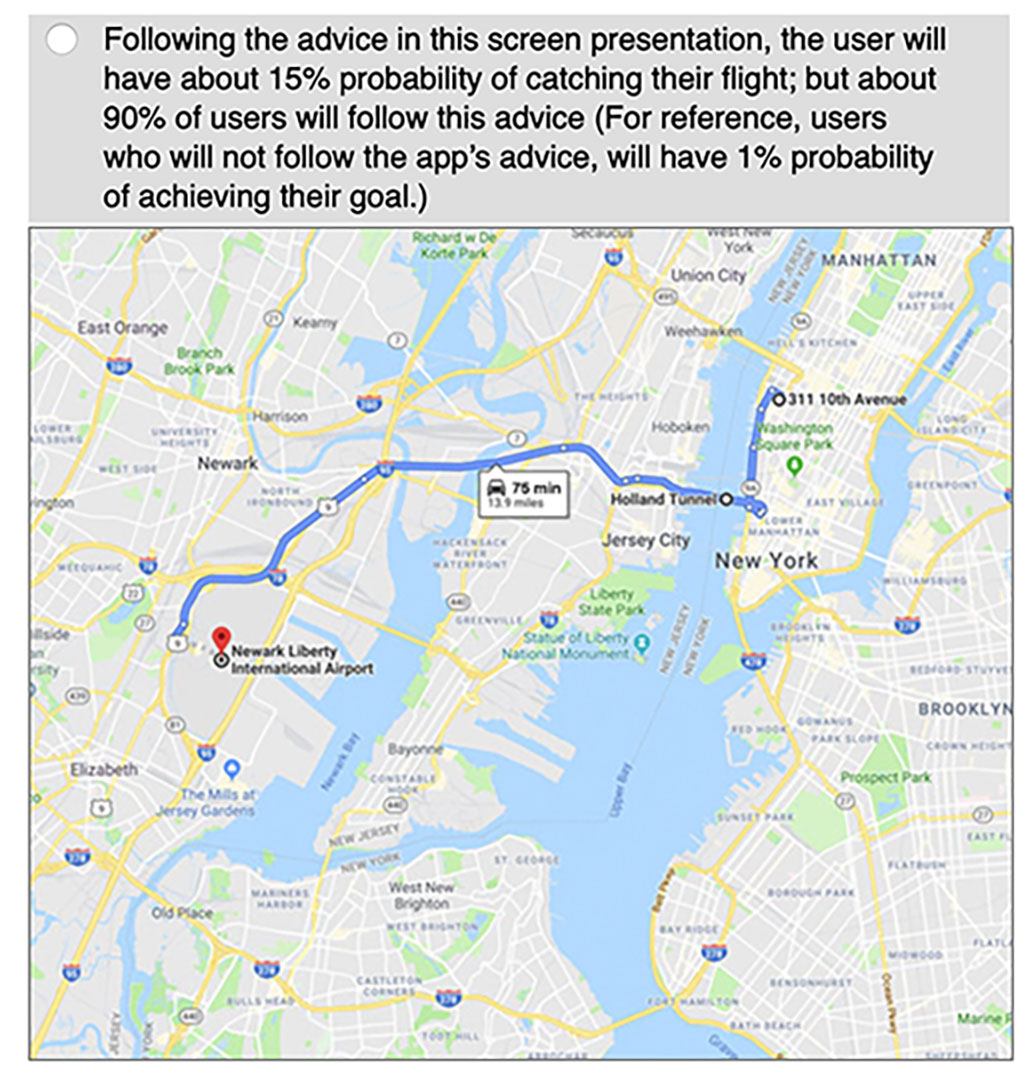}
  \caption{`Adoption-Directed' advice: Participants are shown a route that is shorter in distance but longer in duration. They are instructed that users will have a 15\% probability of catching their flight, but that 90\% are likely to adopt it.}
  \label{fig:second_scenario}
  \Description{Screenshot of the `Route planning' scenario, showing `Adoption-directed' advice.}
\end{subfigure}

\caption{Screen designs from the `Route planning' scenario, where the goal-adoption gap is 90-15 and the `goal' term is given first in the description text. We show, (a) Goal-Directed advice and (b) Adoption-Directed advice. The text above each image explains the advice-type choices to participants.}
\label{fig:image1}
\end{figure}

To address RQ1, we selected four different domain scenarios representing different contexts where users might seek advice.  We designed each scenario to present a situation in which the advice most optimal for individual goal attainment is likely to appear counter-intuitive or seem challenging to adhere to. In these scenarios, we selected the details of both advice types based on realistic examples of advice offered in similar situations (sources referenced in the descriptions below). These scenarios were:

\begin{itemize}
    
    \item \textit{Route planning:} In this scenario (Figure ~\ref{fig:scenario1}), participants were asked to consider the case of a driver aiming to catch a flight and using an app to direct them to an airport in an unfamiliar location. It was based on examples of similar route planning (e.g. using Google maps) under different traffic conditions.

    \item \textit{Investment planning:} This scenario described a novice investor planning for retirement and aiming to recover from the previous year's poorly performing stock market. Two investment plans were shown. It was based on previous research into investment planning for retirement \cite{gunaratne2015informing,gunaratne2015influencing}.

    \item \textit{Training for a 5k run:} The third scenario presented a first time runner using a fitness-training app to guide their preparation for a 5K race in which they aim to finish in less than 26 minutes. It was based on training advice from \cite{gallwayrun,lacke2017run}.

    \item \textit{Making healthier lifestyle choices:} The fourth scenario described a user planning healthier lifestyle choices following a medical checkup. It was based on exercise and diet guidelines from \cite{usda2015diet}.
    
\end{itemize}

We selected four different domains in order to test whether participants' choices are sensitive to domain. The domains selected represent typical areas where algorithmic advice might be offered, but each has particular characteristics that make them interestingly different. For example, we selected `route planning' because the goal is immediate, whereas `training for a 5k run' offers a more long-term but well defined goal. In the `making healthier lifestyle choices' scenario the goal is both longer-term and less concrete, while we selected an `investment planning' scenario were the goal is in the longer-term future but decisions are made in response to historic activity. In each of these scenarios, participants were presented with a choice of two screen designs (e.g. Figure ~\ref{fig:image1} ). One screen shows Goal-Directed advice that is either counter-intuitive (e.g. in the route planning scenario the driver is required to double back and drive a significantly longer distance) or challenging to adhere to (e.g. in the making healthy lifestyle choices scenario the plan set strict rules for diet and exercise).  Participants were told that fewer people were likely to follow the advice presented on this screen, but that if they did manage to adhere to it, they would have a greater chance of achieving their goal. On the second screen design we presented Adoption-Directed advice that might appear more immediately intuitive (e.g. in the route planning scenario the route was shorter in distance and appeared more direct) or easier to adhere to (e.g. in the making healthier lifestyle choices scenario the plan set looser, less stringent, targets). Participants were told that more people were likely to follow this advice, but that they would have a reduced chance of completely achieving their goal.

\subsection{Participant Roles}
To address RQ2, and investigate whether preferences would manifest differently depending on whether participants were offering or receiving advice, we randomly assigned each participant to one of three roles, \textit{developer}, \textit{user}, and an \textit{ethical choice} role. By assigning participants the role of developer we placed them in a scenario where information about likely rates of goal success and adoption would be available, therefore highlighting the dilemma as it may manifest in realistic settings. These participants were asked, ``If you were the app developer, what would you show to the user in this situation?''. In assigning participants the role of user they were placed in a setting akin to a user study where they were presented with goal success and adoption information to elicit a hypothetical choice. These participants were asked, ``If you were the app user, what would you want to be shown in this situation?''. By assigning participants the ethical choice role we were presenting a scenario where they would need to know about the potential for individual goal success and likely adoption rates in order to weigh up the ethics of their advice type preference. These participants were asked, ``Which version of the screen below is more ethical advice to give to the user?''. We did not provide a definition of what `ethical' should be, rather preferring to leave this to participants' own conceptualization of the term. We did this in order to allow participants to focus on responding to the dilemma posed, rather than potentially taking issue with an externally imposed understanding of what is ethical. Our advice choice rationale question, discussed below, allowed us the opportunity to unpick possible issues that might arise from participants' different conceptualizations. 

\subsection{Additional Study Design Considerations}
\subsubsection{Goal-adoption gap}
We also wanted to account for the possibility that participants' preferences would be influenced by the apparent size of the gap between the probability that users would achieve their goal and the probability that users would follow the advice. To control for this, we decided to create two variations and randomly assign participants to one of these. In the first, participants selected between a Goal-Directed advice option with a 90\% goal attainment probability but 15\% adoption rate, and an Adoption-Directed option with 15\% goal attainment probability but 90\% adoption rate. In the second variation, the Goal-Directed advice option presented 60\% goal attainment probability but 45\% adoption rate, and the Adoption-Directed option presented 45\% goal attainment probability but 60\% adoption rate. In all cases participants were told that choosing to follow \textit{neither} advice would result in the user having only a 1\% probability of achieving the goal. This was to highlight to participants that both advice options would be valuable to the user.

\subsubsection{Ordering effects of interface elements}
To control for possible ordering effects, we randomly varied the presentation of advice type options on the page, such that half the time Goal-Directed advice was on the left-hand side and Adoption-Directed advice was on the right, and the other half this was reversed. We also randomly varied the order of the terms referencing Goal-Directed and Adoption-directed advice in the description text above the screen designs. Half the time, the Goal-Directed probability was presented first, e.g. in the route planning scenario with a 90-15 goal-adoption gap the descriptions would read: ``Following the advice in this screen presentation, the user will have about 90\% probability of catching their flight; but about 15\% of users will follow this advice''. For an example see Figure ~\ref{fig:image1}. In the remainder, the Adoption-Directed probability was presented first, e.g. in the investment scenario with a 60-45 goal-adoption gap the descriptions would read: "About 60\% of users will follow this advice; but following the advice in this screen presentation, the user will have about 45\% probability of reaching their retirement goal".   
\subsubsection{Attention test} 
To filter out participants who simply rushed through the study without taking time to consider their responses, we also included a simple attention test question on the main study page based on the content of the particular scenario being shown. This type of test is commonly added to experimental tasks undertaken on MTurk.

\begin{figure}[t]
 \centering
  \includegraphics[width=0.9\linewidth]{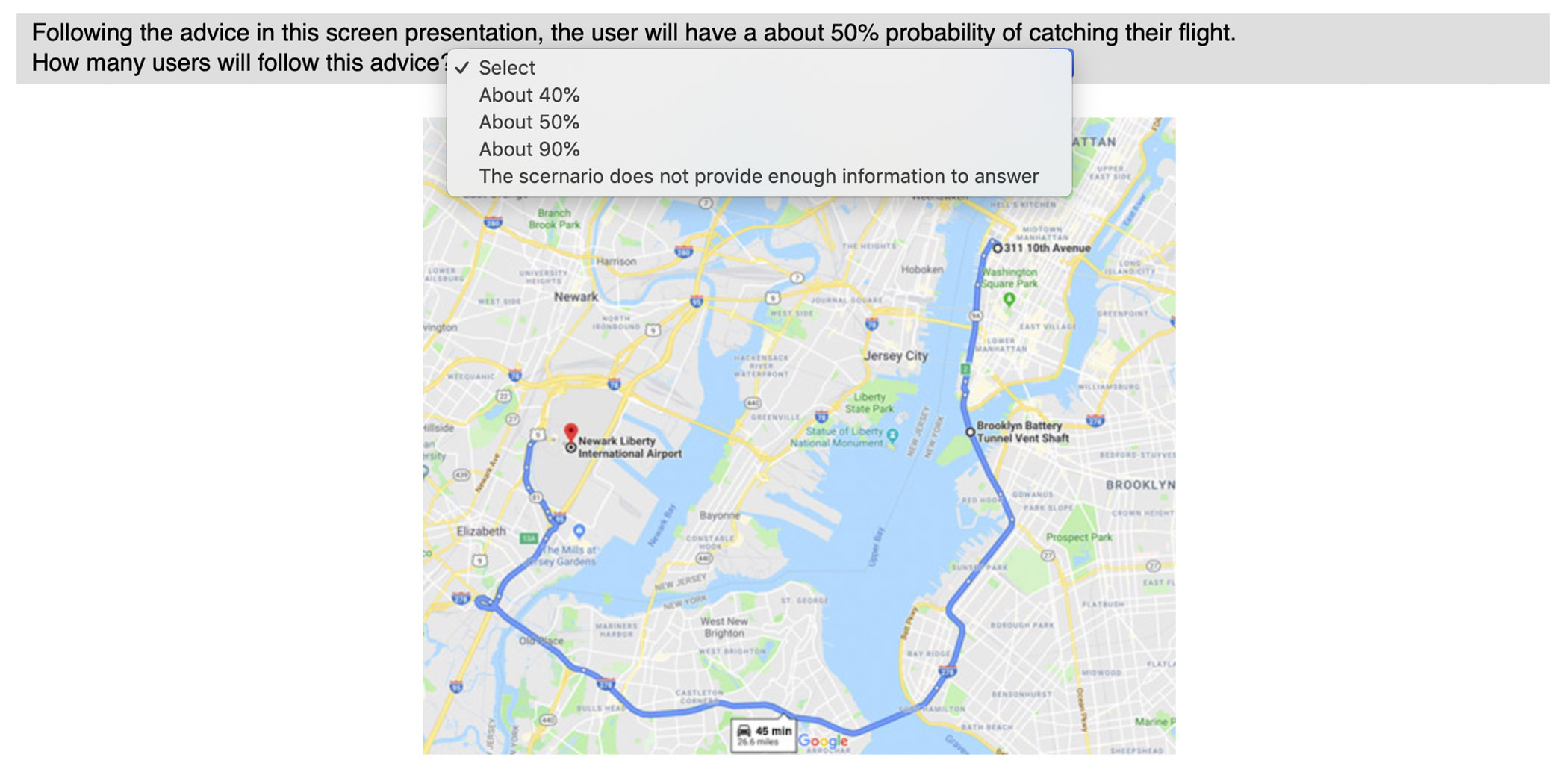}
  \caption{Screenshot of the second scenario test of participants' understanding of independence between the goal attainment and adoption rate probabilities in the `Route planning' scenario.}
  \label{fig:second_scenario}
  \Description{Screenshot of scenario 2 the question testing understanding of the independence of goal attainment and adoption rate probabilities, in the `Route planning' scenario.}
\label{fig:image2}
\end{figure}

\subsubsection{Independence of goal attainment and adoption rate probabilities}
Because we were asking participants to make a design decision regarding which advice type to present to users, rather than which advice they would follow themselves, we wanted to be confident that participants understood the trade-off the dilemma poses. One important aspect of this was understanding that the predicted goal attainment and likely adoption rate probabilities were independent of each other, rather than on an interconnected `sliding-scale', such that an increase in one would automatically lead to an equivalent reduction in the other. To test for this, we added a multiple choice question set in a different domain scenario than the participant had been presented in the main study question. Here we presented participants with a single screen design, saying that it would offer users a 50\% probability of achieving their goal. We then asked them to select from 4 choices the proportion of users who would likely follow this advice (the correct response being: `The scenario does not provide enough information to answer', see Figure ~\ref{fig:second_scenario}).

\subsection{Advice choice rationale}
To better understand participants choice between advice types, and to probe the reasons behind these preferences, we also asked them to briefly explain their choice. First we reminded participants of the selection they had made in the main scenario, and then we asked them for a brief free-text response explaining why they selected their chosen advice type. We also probed participants about their responses to the second scenario in a similar fashion, using these responses to sanity check our decision to test for an understanding of goal attainment and adoption rate independence.

\section{Analysis}
In order to have more confidence in our findings, we adopted a conservative approach to data inclusion. First we discounted 3 submissions from participants who did not complete the task or who spent less than one minute on the task, as we considered these unreliable. We then discounted data from 434 participants who responded incorrectly to the attention check question, the answer to which could be readily found in the text describing the scenario. Such attention tests are commonly used to address potential concerns over data validity in studies where participants are recruited through MTurk   \cite{hauser2016attentive,landers2015inconvenient}, and this is not an atypical number, as up to 42\% of participants in MTurk studies have been found to be inattentive \cite{fleischer2015inattentive}. We then also removed data from 444 participants who responded incorrectly when tested about the independence of advice type probabilities. Data from these participants was discounted because we felt that a failure to understand the independence of the probabilities shown for Goal-Directed and Adoption-Directed advice indicated that participants may not clearly understand the choices available when the dilemma presented. This is a relatively large number of participants to exclude, and we speak further to this decision in the Discussion section. After the removal of data from 772 of 1,589 candidates, we were left with a final data set of 817 participants.

For our quantitative analysis, we first separated participants' response data according to their role assignment, i.e. user, developer, or ethical choice; and the scenario they were assigned, i.e. `route planning', `investment planning', `training for a 5k run', or `making healthier lifestyle choices'. The number of participants in each category is recorded in Table \ref{table:categories}.

\begin{table}
\centering
\begin{tabular}{|c | c c c c|} 
 \hline
  & Scenario & & & \\
 \hline
 Role & Healthy Living & Investment Planning & Race Training & Route Planning\\
 \hline
 developer & 68 & 58 & 76 & 88\\
 ethical & 74 & 73 & 55 & 81 \\
 user & 54 & 55 & 53 & 82\\
 \hline
\end{tabular}
\caption{Number of participants in each experimental category.}
\label{table:categories}
\end{table}

We then compared the rate at which participants in each sub-group preferred the Goal-Directed advice to a 0.5 `indifference' threshold, using one-sample z-tests. A 0.5 reference threshold was used for comparison because we would expect participants to be \textit{indifferent} if 50\% of the sample selected the goal-directed advice and 50\% selected the adoption-directed advice. This initial analysis provides us with a high level view of the preferences of each of these sub-groups. 

To extend our analysis, we used a logistic regression to investigate how these variables might influence an individual's probability of selecting the Goal-Directed advice. To do this we created three models. In Model 1 we included the participant's assigned role and scenario as main features in the regression. For Model 2 we also included terms for covariates: the goal-adoption gap, the side of the page on which the advice types were presented, the order that advice types were listed in the description text, participant gender, the log of participant age, and the log of time taken to complete the study. Finally, in Model 3, we included an interaction term between participant role and scenario. We use logistic regression because our dependent variable and the majority of our independent variables were categorical.

We also analyzed these participants' responses to the post-study question probing their advice choice rationale. We received responses from 814 of the 817 participants included in our quantitative analysis. These responses were brief, typically somewhere between a few words and a short paragraph in length, with a mean response length of 24 words. There was no need to sanitize these responses, most likely due to conservative approach we took to data inclusion (described above). We performed a simple thematic analysis \cite{braun2006using}, which involved an initial close reading and clustering led by the first author, followed by refinement and agreement seeking with one other researcher. For a first pass, the responses from all participants were considered together, resulting in a high-level set of main themes. Following this, we divided the responses into two groups so that those who selected a preference for Goal-Directed advice and those who selected Adoption-Directed advice were considered independently. This provided us with a more fine-grained understanding of the differences between the two groups and similarities within them. Having identified patterns across the data, and extracted key ideas, we sought agreement across interpretations, through discussion and where necessary testing alternative framings.

\begin{figure}[b]
 \centering
  \begin{subfigure}{0.49\textwidth}
  \captionsetup{margin=0.1cm}
  \includegraphics[width=0.9\linewidth]{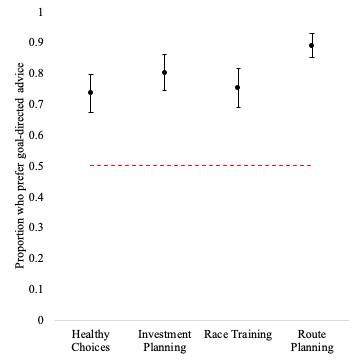} 
  \caption{Preference by Scenario}
  \label{fig:scenario_graph}
  \Description{Graph showing the proportion of people preferring Goal-Directed advice according to the scenario presented, with 95\% confidence intervals.}
  \end{subfigure}
  \begin{subfigure}{0.49\textwidth}
  \captionsetup{margin=0.1cm}
  \includegraphics[width=0.9\linewidth]{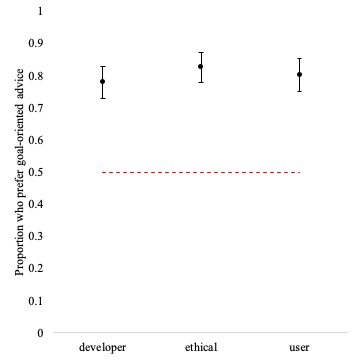}
  \caption{Preference by Role}
  \label{fig:role_graph}
  \Description{Graph showing the proportion of people preferring Goal-Directed advice according to the role assigned, with 95\% confidence intervals.}
 \end{subfigure}
 
\caption{Proportion of people preferring Goal-Directed advice according to: (a) the scenario presented; and (b) the role they were assigned, with 95\% confidence intervals. Red lines represent the 0.5 `indifference' threshold}
\label{fig:image3}
\end{figure}

\begin{table} [t]
  \caption{Results of logistic regression testing effect of role and domain on probability of selecting Goal-Directed advice. Model 1 is the model of main effects; Model 2 includes covariates; Model 3 includes an interaction effect between role and domain.}
  \label{tab:log_reg}
  \begin{tabular}{lllll}
    \toprule
     & &Model 1&Model 2&Model 3\\
    \midrule
    Intercept & & 2.07(0.26)***&3.26(1.44)* & 3.99(1.51)**\\
    Role & \textit{developer} & -0.08(0.22) &-0.07(0.22) & -0.54(0.54)\\
    &\textit{ethical} & 0.22(0.22)& 0.24(0.23)& -0.46(0.56)\\
    Domain & \textit{health} & -1.11(0.26)***& -1.11(0.27)***& -1.45(0.53)**\\
    &\textit{investment}& -0.75(0.28)**& -0.75(0.28)**&-1.45(0.53)** \\
    &\textit{race}& -1.08(0.27)***& -1.08(0.27)***& -1.51(0.53)**\\
    Gap size & \textit{90-15 split} & & 0.11 (0.18)& 0.10(0.18)\\
    Text order & \textit{goal first, adoption second} & & 0.09(0.18)& 0.07(0.18)\\
    Screen side & \textit{goal left, adoption right} & & -0.09(0.18)& -0.10(0.18)\\
    Gender & \textit{male} & & -0.21(0.18)& -0.19(0.19)\\
    &\textit{other} & & -0.04(1.15)& -0.05(1.15)\\
    Age & & & -0.41(0.31)& -0.45(0.31)\\
    Time on task& & & 0.07(0.19)& 0.03(0.19)\\
    Role*Domain&\textit{developer*health} & & & 0.42(0.68)\\
    &\textit{developer*invest.} & & & 1.16(0.72)\\
    &\textit{developer*race} & & & 0.31(0.67)\\
    &\textit{ethical*health} & & & 0.60(0.70)\\
    &\textit{ethical*invest.} & & & 0.94(0.71)\\
    &\textit{ethical*race} & & & 1.11(0.73)\\
    \midrule
    AIC & & 793.08&801.59 &808.00 \\
    Pseudo R$^2$ & &0.033 &0.037 &0.044\\
  \bottomrule
  \multicolumn{5}{l}{(Signif. codes:  $<0.001 "***"; 0.001 - 0.01 "**"; 0.01 - 0.005 "*"; 0.05 - 0.1 "^t" $)} \\
\end{tabular}
\end{table}

\section{Findings}
Of the 817 participants in our final data set, 658 (80.54\%) selected a preference for displaying Goal-Directed advice over Adoption-Directed advice (z=17.46, p<0.05, 95\% C.I.[0.78,0.83]). This preference was consistent across all scenarios and all roles, and in all cases this preference was statistically significant using a one-sample z-test of proportions relative to the 50\% indifference threshold. With regard to the different domain scenarios, we found that 74.49\% of participants shown the `making healthier lifestyle choices' scenario (z=8.31,p<0.05,95\% C.I.[0.70,0.79]), 80.64\% of participants shown the `investment planning' scenario (z=10.72, p<0.05, 95\% C.I. [0.77,0.84]), 74.45\% of participants shown the `training for a 5k run' scenario (z=8.24, p<0.05, 95\% C.I. [0.70,0.79]), and 89.64\% of participants shown the `route planning' scenario selected a preference for Goal-Directed advice (z=15.43, p<0.05, 95\% C.I. [0.87,0.92]). These rates are visualized in Figure~\ref{fig:scenario_graph}. Similarly, 78.28\% of participants assigned the role of developer (z=11.69, p<0.05, 95\% C.I. [0.75,0.82]), 83.04\% of those assigned the ethical choice role (z=13.97, p<0.05, 95\% C.I. [0.80,0.86]), and 80.32\% of those assigned the role of user (z=11.87, p<0.05, 95\% C.I. [0.77, 0.84]) selected a preference for Goal-Directed advice (Figure~\ref{fig:role_graph}).

Extending beyond these high level findings, the results of the logistic regression (Table~\ref{tab:log_reg}) indicates that the scenario participants were shown significantly influences the probability that they select a preference for Goal-Directed advice. In particular, Models 1 and 2 show that participants shown the `route planning' scenario were significantly more likely to select this preference  relative to the other three scenarios. By converting the coefficients on scenario from Model 2 to probabilities (holding age and time on task at their medians - 35 years and 312 seconds, respectively), we see that the predicted probability of an individual selecting Goal-Directed advice is 94.23\% when shown the `route planning' scenario. This is significantly higher relative to the other scenarios: 84.39\% in the `making healthier lifestyle choices' scenario ($\beta$ = -1.11, S.E. = 0.27, z=-4.14, p<0.000, 95\% C.I. [-1.63,-0.58]), 88.55\% in the `investment planning' scenario ($\beta$=-0.75, S.E. = 0.28, z = -2.67, p=0.008, 95\% C.I. [-1.30,-0.20]), and 84.79\% in the `training for a 5k run' scenario ($\beta$=-1.08, S.E. = 0.27, z = -3.99, p=0.0001, 95\% C.I. [-1.60,-0.55]). These comparisons are significant even when correcting for multiple comparisons using a Bonferroni correction with a threshold of p=0.017. On the other hand, Model 3 shows that participants' did not differ in their preference for Goal-Directed advice across roles.

Findings from our analysis of participants' responses to the post-study choice rationale question, explaining \textit{why} they selected a preference for presenting users with Goal-Directed advice or Adoption-Directed advice, offer further insight into these quantitative results. A number of examples from this analysis are included in the discussion that follows, here we briefly list the themes that emerged. 

\begin{itemize}
    \item \textit{Relationships between individual users and the crowd:} Some participants selecting a preference for Goal-Directed advice explained their choice in terms of \textit{supporting an individual user in successfully achieving their goal}. Others stressed that it was \textit{the responsibility of individuals whether they choose to follow advice or not}, or \textit{wanted to reward those with the capacity to adhere to advice}, or adopted an \textit{egocentric standpoint} stressing their own qualities in comparison to others. For participants who selected Adoption-Directed advice, \textit{concern for the wider benefits of a greater number of people} or the \textit{pragmatic partial achievement of a goal by larger numbers of users} were important explanations.
    \item \textit{Personal insight and domain experience:} This was a strong theme among participants selecting Adoption-Directed advice, who made \textit{decisions based on heuristics} such as `diversify investment'. Those selecting Goal-Directed advice often made \textit{connections to their own lifestyles or practices}.
    \item \textit{Information detail and complexity:} For those selecting Goal-Directed advice, the \textit{discipline imposed by a more detailed plan} would make it easier to follow through to successful goal achievement. Participants selecting a preference for Adoption-Directed advice suggested the less detailed and demanding option would be \textit{a more realistic plan}.
    \item \textit{Success of the app or company:} Typically found among participants in the developer role, but across all scenarios and both advice types, this theme stressed success of the app, or the company behind it. For participants selecting Adoption-Directed advice the focus was on attracting a high number of downloads and \textit{large initial user base}, so that the app would be a viable business proposition. For participants selecting Goal-Directed advice the focus was more on \textit{user retention and satisfaction}, on minimizing bad reviews and increasing good ones. 
\end{itemize}

\section{Discussion}
Our findings show participants indicating a clear preference for offering users Goal-Directed over Adoption-Directed advice, in the tasks we presented. This was consistent, albeit with varying rates of preference, across each of the four scenarios we tested in order to address RQ1, and was also invariant to the role participants were assigned in order to address RQ2.   This suggests that most participants focused on those users able to adopt and then adhere to the advice being offered, rather than the wider population. Such a finding is consistent with prior findings about  moral dilemmas, (e.g. \cite{bonnefon2016social,thomson1976killing}), which suggest that people generally avoid decisions that are sub-optimal at the level of individual utility, even if they increase aggregated social utility. Our  findings about participants' choices are echoed in our analysis of participants' responses to the advice choice rationale question, where one of the themes to emerge under the theme \textit{`Relationships between individual users and the crowd'} was \textit{`supporting an individual user in successfully achieving their goal'}, while another emphasized that it is \textit{`the responsibility of individuals whether they choose to follow advice or not'}, e.g.:

\begin{displayquote}    
``The point of investment advice is to achieve the goal - it is up to the individual to decide if they will follow the advice'' (\textit{Investment planning, user, 90-15, Goal-Directed advice})
\end{displayquote}

\begin{displayquote}
``The app can't control who chooses to follow their advice, but they can control the quality of that advice, so I felt it was more ethical to provide investment advice that had a higher likelihood of helping a committed client reach his/her investment goal.'' (\textit{Investment planning, ethical, 90-15, Goal-Directed advice})
\end{displayquote}

Such views were not universal though. A significant number of participants selected Adoption-Directed advice, and the rationale question responses indicate that \textit{`concern for the wider benefits of a greater number of people'}, which might accrue through the \textit{`pragmatic partial achievement of a goal by a large number of people'}, were key motivators. This was fairly common among participants presented with the `investment planning' and `making healthy lifestyle choices' scenarios, but rarer among those shown the `training for a 5K run' scenario; and there were no examples at all amongst those in the `route planning' scenario. Example responses include:

\begin{displayquote}
``Getting more people to consider their investment goal is still very beneficial, even if they fall short of their goal.  This scenario meant that more people would be thinking about and taking action with their investments.'' (\textit{Investment planning, developer, 60-45, Adoption-Directed advice})
\end{displayquote}

\begin{displayquote}    
``I think it would help more people make some steps toward a healthier lifestyle which is better to me than fewer people achieving completely healthy lifestyles.'' (\textit{Making healthy lifestyle choices, developer, 60-45, Adoption-Directed advice})
\end{displayquote}

These different preference explanations are one indicator of the influence scenario may play in the selection of advice types; suggesting greater nuance underneath the across the board preference we found in response to RQ2. Another is our quantitative finding that participants shown the `route planning' scenario were significantly more likely to select a preference for Goal-Directed over Adoption-Directed advice. Each of these points to the importance of task complexity and task familiarity in the decisions participants made.

\subsection{Task complexity}
The `route planning' scenario is arguably simplest, and most likely to manifest in the wild as a one-shot binary-selection. In this scenario, success only benefits the individual user, and there can be no partial achievement of the goal. The user either arrives in time to catch their flight, or they do not. Because of this, research showing the importance of loss aversion in decision-making e.g.\cite{kahneman979prospect,barberis2013thirty} might offer an explanation for our finding that participants shown this scenario were significantly more likely to prefer Goal-Directed over Adoption-Directed advice, when compared with participants shown any one of the other three scenarios. In contrast, the `making healthier lifestyle choices' scenario is likely the most complex when translated into real world experiences. Yet here too task complexity played a role. For participants preferring to offer Goal-Directed advice, the \textit{`discipline imposed by a more detailed plan'}, and the specific targets set would make it easier to follow through to successful goal achievement, e.g.:
\begin{displayquote}
``The instructions were much more rigid and difficult to achieve. For example, where one plan said to simply reduce alcohol consumption, the other listed an exact number to stay below. I felt that people that followed the more detailed plan received a better outcome than following a very basic and vague outline of a plan.'' (\textit{Making healthy lifestyle choices, user, 90-15, Goal-Directed advice})
\end{displayquote}

However, participants shown this scenario but preferring  Adoption-Directed advice suggested the less detailed and less demanding option would be simpler to understand and \textit{`a more realistic plan'}, that would also be easier to adhere to. This supports research suggesting users are put-off following advice from online recommender systems if it is considered demanding or challenging, e.g. \cite{schwind2011will}.
\begin{displayquote}
``Based on what I've seen, most people would take one look at the other plan and run away from it.  There's too much to it and people don't want to be bothered with programs like that.  I think more people would be willing to take the common sense approach, which is what the plan I chose was.'' (\textit{Making healthy lifestyle choices, developer, 60-45, Adoption-Directed advice})
\end{displayquote}

Further research is needed to better understand the impact of task complexity on participants' preferences. In this study, the choice between Goal-Directed and Adoption-Directed advice was presented as a binary selection between static screen designs. While this was intentional, and allowed us to isolate particular issues of concern, additional research is needed to fully investigate the nuances within more complex evolving interactions. Future studies should consider a longitudinal approach, that allows a deeper dive to study of how different options and trade-offs might be captured and expressed dynamically for users to explore and select between.

\subsection{Task familiarity}
In addition to being simplest, the `route planning' scenario may also be the most familiar to participants. The use of journey planners is widespread, and research indicates that in circumstances where travel time is the key factor users are likely to follow their advice \cite{samson2019exploring}. For the large number of participants who selected a preference for Goal-Directed advice when presented with this scenario, it may be that trust in this technology, and familiarity with the visualizations associated with it, reduces the apparent counter-intuitiveness of the advice offered. Further evidence of the impact that task or scenario familiarity might have on participants' advice type preferences comes via references to \textit{`personal insight and domain experience'} in their explanations of advice type selections. This was evident among participants that selected Adoption-Directed advice who often made \textit{`decisions based on heuristics'}, for example: 
\begin{displayquote}
``Because historically, placing all your eggs in one basket is bad.  Better to spread the risk and be more diversified.  A fund that does well in the previous year will not necessarily do well in the future.'' (\textit{Investment planning, ethical, 90-15, Adoption-Directed advice})
\end{displayquote}

While those who selected Goal-Directed advice often  made \textit{`connections to their own lifestyles or practices'}, such as:

\begin{displayquote}    
``I don't mind taking back roads or scenic detours as long as I get to my destination on time.  I know big city traffic can be horrible especially during rush hours and on popular roads.  Sometimes taking a round about route gets you to your destination faster.'' (\textit{Route planning, user, 90-15, Goal-Directed advice})
\end{displayquote}

We suspect that participants make these personal references for reasons closely related to those underlying research that suggests decision-makers privilege self-related information and adopt an egocentric standpoint, e.g. \cite{kruger1999lake,krueger2001role,clement2000primacy,dunning1996evidence}, and to research suggesting decision-makers may discount advice because they have access to their own justifications, but not to those of others \cite{yaniv2000advice}. Because of this, future studies might investigate how advice types can be personalized.

\subsection{Cultural differences}
Participants' average preference for Goal-Directed over Adoption-Directed advice may also reflect dominant attitudes towards individual achievement versus collective benefit in the society our sample is drawn from. We recruited MTurk workers with an account in the U.S, a society where individualism is highly valued \cite{markus1991culture,triandis1990multimethod}. It remains an open question whether our findings would translate to participants recruited from societies considered to place greater value on shared responsibility \cite{hacker2012shared} or collectivism \cite{markus1991culture,triandis1990multimethod}. In addition to potential biases associated with national cultures, there may also be a bias in the culture of participation in MTurk. Paid crowd workers may make different choices from participants recruited in other ways. Further research should investigate this, as potentially analogous behavior has been seen in comparisons between tasks undertaken by crowd workers and citizen scientists \cite{cartwright2019crowdsourcing}.

This potential bias could  explain the many examples in which participants who preferenced Goal-Directed advice adopted an \textit{`egocentric standpoint'} in their explanation that was explicitly made in opposition to others, or which stressed the individual qualities of the participant themselves, e.g.:

\begin{displayquote}
``I'm after the results. I don't care whether others are willing to put in the effort - I am.'' (\textit{Making healthy lifestyle choices, ethical, 60-45, Goal-Directed advice})
\end{displayquote}

\begin{displayquote}
``I don't need to feel validated by others' choices.  If I know I have a higher chance of reaching my goal I'm going to take it.'' (\textit{Route planning, user, 90-15})
\end{displayquote}

\begin{displayquote}
``I am a goal oriented person, once I start a goal I almost ALWAYS reach it. So for me, that scenario was more appealing.'' (\textit{Training for a 5k run, developer, 90-15, Goal-Directed advice})
\end{displayquote}

However, statements such as these may also reflect previous research which suggests decision-makers make overconfident self-assessments \cite{block1991overconfidence} and overestimate their own abilities \cite{kruger1999lake}. It could be that participants who are able to simply `make a change', e.g. when dieting, show a preference for Goal-Directed advice and simply assume that others are similar. Further research is needed to unpack the more nuanced impact of culture on peoples' advice type preferences, as design decisions based on social cue data gathered in a single country, or from within a single online community, may not translate into other contexts. 

\subsection{Design values and ethical considerations}
In making decisions about what type of advice is best to offer users, designers are asked to make value judgments in complex and ethically ambiguous situations.  These judgements, and the values they operationalize, may be subject to different interpretations when employed in the service of an altruistic cause than when used to drive commerce \cite{chivukula2019analyzing}. Yet, despite the influence designers have on peoples' lives \cite{friedman2003human,redstrom2006persuasive,buchanan1985declaration}, prior discussion of the role of ethics in the design of technologies that offer advice or try to persuade users has tended to focus on the role they might play in the physical \cite{purpura2011fit4life}, material \cite{gunaratne2015influencing}, and psychological welfare \cite{kehr2012transformational} of individuals, considering societal good as reflecting value judgments on what might be considered `good' for the individual user, e.g. \cite{berdichevsky1999toward}, rather than possible wider impacts. Excepting `route planning', in each of the scenarios we presented there are potential benefits for users who partially achieve their goal. These benefits would be additional to the benefit gained by those who ultimately achieve their goal in full. The ambiguity surrounding such judgments is also reflected in themes from our analysis of the choice rationale responses. This highlights important ethical questions for designers who may have to choose between showing people what they apparently want as individuals, versus showing them what may have a larger positive impact on the greatest number of people. This contrast, between the individual and the group, adds an additional layer of complexity to decisions about `good for whom', in a similar way to how feminist HCI \cite{bardzell2011towards} and HCI for sustainability \cite{blevis2007sustainable,disalvo2010mapping} have called on designers to reconsider what it means to `do good'. 

Related to the ethical considerations of `good for whom' are questions about `good for business'. Among participants assigned the developer role there was a repeating pattern of explanations that considered the \textit{`success of the app or company'}, to be the primary motivation for their selection of advice type to offer. This theme crossed all scenarios and both advice types, with subtle differences. For participants selecting Adoption-Directed advice the focus was on attracting a high number of downloads and \textit{`large initial user base'}, so that the app would quickly become a viable business proposition. For participants selecting Goal-Directed advice the focus was more on \textit{`user retention and satisfaction'}, on minimizing bad reviews and increasing good ones, so that the business was viable in a more ongoing sense. Examples include:

\begin{displayquote}    
``If you are looking to generate money and a large user base with the app, then I think you would want to drive more traffic to your app.  In that case, you want the option where 90\% of people would participate.'' (\textit{Making healthy lifestyle choices, developer, 90-15, Adoption-Directed advice})
\end{displayquote}

 \begin{displayquote}   
``Because I'd want the app to earn a good reputation by giving correct advice'' (\textit{Route planning, developer, 90-15, Goal-Directed advice})
\end{displayquote}

\subsection{Experimental design choices}
The choices made in designing any study are reflected in its findings. In this study, the choice between Goal-Directed and Adoption-Directed advice was intentionally presented as a one-shot binary choice, following a convention familiar in choice experiments undertaken by economists, e.g \cite{bansback2012using,egesdal2015estimating,horne2005improving}. We also asked participants to adopt one of three different roles when considering their selection of Goal-Directed or Adoption-Directed advice, and we selected four scenarios in which to probe participants' preferences. While similar techniques have been widely used in prior research, we should not exclude the possibility that differences observed are a function of the experiment itself. In order to gain more nuance to our understanding, and to more properly distinguish between possible experimental effects and deeper preferences, future research should consider longitudinal studies and qualitative methods. This would support investigation into if and how design choices can influence adoption rates for counter-intuitive or challenging advice, and therefore reduce the dilemma's impact. These might be considered in the context of health, where relationships with advice givers are likely to be ongoing, and where short and long-term goals interact. To better understand the dilemma, future research should also consider including additional conditions in which Goal-Directed and Adoption-Directed advice align. Adding conditions in which the probability of individual goal attainment and the probability that the advice will be adopted are both high, and conversely where they are both low, will help to further pick apart peoples' preferences.

\subsubsection{Testing for understanding the independence of advice type probabilities}
To be confident that we only included data from participants who understood the trade-offs that the dilemma poses, we included a test in a second domain. Failure of this test excluded the responses of 444 participants. While we consider this the correct approach for making an initial study into the advice dilemma we identify, we also acknowledge the impact of this choice. For this reason we compare the advice type preferences of this set of 444 participants with those of the 817 participants included in our full analysis. In this comparison, we first see an overall reduction in the preference for Goal-Directed advice over Adoption-Directed advice: 65.9\% (290/444) against 80.5\% (658/817). A chi-square test of independence between these two samples shows that these differences are significant ($\chi^{2}$=32.23, $p$<0.00). When broken down by domain scenario, participants in each group also indicated a reduced preference towards the Goal-Directed advice. In three of the four scenarios this difference was significant.  In the `investment planning' scenario the difference was 60.83 \% against 80.64\% ($\chi^{2}$ = 13.50, $p$ < 0.00); in the `training for a 5k run' scenario it was 54\% against 74.45\%  ($\chi^{2}$ = 11.40, $p$ < 0.00); and in the `route planning' scenario it was 76.56\% against 89.64\% ($\chi^{2}$=10.5, $p$=0.001).

Despite these differences, our initial results are robust and the findings remain largely the same with or without the exclusion. However, they do indicate that there is likely more nuance to the preferences people might have in actual use than is expressed in this initial experiment. We believe that this indicates the potential importance of numeracy skills on participants' selections, and in understanding the trade-offs the dilemma introduces. It also further indicates that research is needed into the way the dilemma manifests in the wild, where users may be exposed to repeated ongoing advice, making longitudinal decisions where short-term and long-term motivations and goals interact, and where issues of low numeracy and low engagement may be common.

\subsection{Algorithmic Advice Dilemmas: Future Research}
The experimental design of this study, i.e. a one-shot, discreet choice between advice types, was chosen in order to isolate the particular cases where advice might appear counter-intuitive or challenging, and echoes a tradition of similar experimental studies in economics. However, in practice this advice dilemma may also appear in cases where users have more complex, ongoing relationships with applications and recommender systems that involve use cases of repeated or adjusted advice \cite{wang2018tem}, often based on users' dynamic preferences and previous choices \cite{kapoor2015like, moore2013taste}, and other behavioral patterns such as in sequence-aware recommender systems \cite{quadrana2018sequence, quadrana2018sequence2}.

Our quantitative findings show that participants shown the `route planning' scenario, the scenario we consider most likely to be a one-shot decision, were also significantly more likely to select Goal-Directed advice. We also see from participants' response to the choice rationale question, that it was those shown the `making healthier lifestyle choices' scenario who most often mention task complexity, regardless of whether they selected a preference for Goal-Directed advice or Adoption-Directed advice. This is the scenario we consider most open ended and likely to present in a situation of regular ongoing interactions between a user and a system offering algorithmic advice. Taken in combination, these two findings strongly suggest that in the context of recommender systems, responses to the dilemma will need to be more subtle and nuanced. What a user prefers may vary across interactions, and may also vary through the temporal unfolding of interactions. Their preferences before advice is offered may vary from their preferences at the time of offering, they may vary again following success or failure in adhering to that advice, and they may vary yet again following information about the choices of other users. Future research, building on work at the intersection of CSCW and recommender systems \cite{chang2015using, koivunen2019understanding, xu2019think}, would therefore be needed to address the design dilemma outlined here in the context of repeated advice and collaborative filtering \cite{daskalova2018investigating}. For example, should algorithmic tools prioritize a sub-optimal `good' advice over optimal `best' advice if the sub-optimal advice is consistent with the choices of users in the network of the focal user, and therefore more likely to be followed? Such questions should be considered not only in terms of their effectiveness in driving users' choices, but importantly, in relation the ethical aspects of recommender systems and their design (e.g \cite{paraschakis2016recommender, milano2020recommender}).

\subsection{Design Considerations}
In this paper we probe a dilemma for designers of algorithmic advice systems. Here we discuss design considerations for these circumstances. These we consider complementary to existing guidelines, e.g. for  designing social recommender systems \cite{arazy2010theory, harman2014dynamics}.   

Once identified, the dilemma between Goal-Directed and Adoption-Directed advice should be surfaced early on in the design process. Human-centered design methods advocate for iterative cycles of contextual inquiry, prototyping, and evaluation; and at each stage we would argue the need for comparative user studies, both qualitative and quantitative, to explore the dilemma as it plays out in-situ. This would allow not only comparison between different advice types, but also inquiry into specific ways they might be dynamically explored and compared by users; e.g. in situations such as route planning where the alternative recommendations may be presented simultaneously, but where designers must choose which advice to suggest as initially preferred. Our findings suggest not only that preferences may vary according to domain and scenario, but also that motivations underlying these preferences may vary too. 

While our findings may indicate participants' clear preference for Goal-Directed advice, a substantial minority of participants selected a preference for Adoption-Directed advice. In practice, the advice design dilemma is often likely to present in complex situations where users have ongoing longitudinal interactions with applications offering advice. The variation we see in the strength of preference between the different domain scenarios, and the explanations participants gave for making these selections, across and within these scenarios, suggest that designers should investigate adaptive approaches to presenting algorithmic advice. At a simple level, this might involve making advice type a dynamic and/or configurable option, either when the user is completing initial setup or in use on a case by case situational basis. However, designers might also consider options that draw on prior research that inspired this work. For example, if we know that users have a tendency to discount good advice \cite{goodwin1999judgmental,lim1995judgemental} and display overconfidence in their own abilities \cite{block1991overconfidence} because of a bias towards self-related, egocentric information \cite{kruger1999lake,krueger2001role,clement2000primacy,dunning1996evidence}, can the choice of advice type be combined with challenges based on predicted goal success and adoption rates? and can the way this is presented to users guide these users towards more effective decision-making?  In such a case, we can imagine designing an application that uses machine learning to predict the likelihood of the advice dilemma, e.g. in advice relating to lifestyle changes, and which then adapts the advice offered dynamically, based first on data from aggregated prior use and then increasingly integrating this with the users' own responses. We can also imagine a potential user who receives advice, such as to quit smoking or to severely reduce salt or sugar consumption, but discounts the advice because they may be overconfident in their capacity to avoid the consequences, or because they are biased towards their current feelings of wellbeing. In this situation, the application might start - upon user consent - by varying the advice, at times using low adoption rates associated with Goal-Directed advice as an egocentric challenge to the user, and at other times using the less challenging but still beneficial goals associated with Adoption-Directed advice to provide a confidence supporting reward. Over time the application might blend its wider understanding of what represents Goal-Directed or Adoption-Directed advice, which is based on data aggregated from many users, with a more nuanced understanding of what these might mean for the particular user in question, and offer advice that responds to the dynamics of a particular situation in a more fine-grained way.

The example of applications that provide users with healthcare advice illustrates the considerations necessary for design and development in the medical domain. Digital therapeutics systems \cite{nebeker2019building}, for managing chronic diseases such as diabetes, are flourishing, with healthcare clinicians now prescribing apps to patients as supplements to their treatment regimens.  In this rapidly evolving situation, designers play an important role in helping shape the messaging that clinicians leverage through new digital therapeutic interactions.  Given a choice, does a physician prescribe the diabetes app that is most challenging to follow but has the best clinical outcomes when adhered to; or one that is more easily adopted by a larger number of patients, albeit with a lower effect size on those that follow its advice?  What factors should influence this decision?  And how might it be affected by the prescriber's inherent biases?  Is there a role for advanced analytics, based on social cues from historic use, in helping to personalize the advice these recommender systems offer, so that the clinician can promote adoption directed messages in patients who are likely to struggle with the more challenging path and goal directed advice for those who demonstrate a higher probability of completion? For the clinician, these trade-offs are not new. However, sharing the way these choices influence patients with designers or developers is.  This research indicates that new partnerships and collaborations between designers and healthcare clinicians are urgently needed to examine these trade-offs and develop best practices.

\section{Conclusion}
In this paper we have made an initial step towards better understanding the advice design dilemma that arises when advice most likely to maximize an individual's chances of achieving their goal is not the same as advice most likely to maximize the number of people who will adopt it. We studied participants' preferences when selecting between Goal-Directed and Adoption-Directed advice online in an experimental study spanning four domain scenarios, three user types, and two variations in the gap between goal attainment and advice adoption probabilities. We found an average preference for advice that favors individual goal attainment over higher user adoption rates, albeit with some variation across advice domains. The implications of these findings are both practical and theoretical. From a practical perspective, choosing between presenting advice that may be best in terms of individual goal attainment but which will reach few users, or advice that is merely good but will be adopted by many, is likely to be an increasingly common dilemma for designers to face in algorithmic advice design scenarios. From a theoretical perspective, the preferences shown for Goal-Directed advice (i.e. advice best for individual goal attainment that will be adopted by few users), and the explanations revealed in participants' survey comments, echo responses to moral dilemmas, e.g. \cite{bonnefon2016social,thomson1976killing}, where people eschew making decisions that are sub-optimal at the individual level, even if they are optimal at the population level. With the rapid growth in applications driven by algorithmic advice based on the social information derived from aggregated prior use, HCI and CSCW researchers and practitioners are likely to increasingly engage with these fundamental questions. Our findings will help inform this conversation.

%
\begin{acks}
This work was supported by the National Science Foundation award \#1928614.
\end{acks}

%
\bibliographystyle{ACM-Reference-Format}
\bibliography{advice}


\begin{thebibliography}{91}


\ifx \showCODEN    \undefined \def \showCODEN     #1{\unskip}     \fi
\ifx \showDOI      \undefined \def \showDOI       #1{#1}\fi
\ifx \showISBNx    \undefined \def \showISBNx     #1{\unskip}     \fi
\ifx \showISBNxiii \undefined \def \showISBNxiii  #1{\unskip}     \fi
\ifx \showISSN     \undefined \def \showISSN      #1{\unskip}     \fi
\ifx \showLCCN     \undefined \def \showLCCN      #1{\unskip}     \fi
\ifx \shownote     \undefined \def \shownote      #1{#1}          \fi
\ifx \showarticletitle \undefined \def \showarticletitle #1{#1}   \fi
\ifx \showURL      \undefined \def \showURL       {\relax}        \fi
\providecommand\bibfield[2]{#2}
\providecommand\bibinfo[2]{#2}
\providecommand\natexlab[1]{#1}
\providecommand\showeprint[2][]{arXiv:#2}

\bibitem[\protect\citeauthoryear{Adomavicius, Bockstedt, Curley, and
  Zhang}{Adomavicius et~al\mbox{.}}{2019}]%
        {adomavicius2019reducing}
\bibfield{author}{\bibinfo{person}{Gediminas Adomavicius},
  \bibinfo{person}{Jesse Bockstedt}, \bibinfo{person}{Shawn Curley}, {and}
  \bibinfo{person}{Jingjng Zhang}.} \bibinfo{year}{2019}\natexlab{}.
\newblock \showarticletitle{Reducing Recommender Systems Biases: An
  Investigation of Rating Display Designs}.
\newblock \bibinfo{journal}{\emph{Forthcoming, MIS Quarterly}}
  (\bibinfo{year}{2019}), \bibinfo{pages}{19--18}.
\newblock


\bibitem[\protect\citeauthoryear{Adomavicius and Tuzhilin}{Adomavicius and
  Tuzhilin}{2005}]%
        {adomavicius2005toward}
\bibfield{author}{\bibinfo{person}{Gediminas Adomavicius} {and}
  \bibinfo{person}{Alexander Tuzhilin}.} \bibinfo{year}{2005}\natexlab{}.
\newblock \showarticletitle{Toward the next generation of recommender systems:
  A survey of the state-of-the-art and possible extensions}.
\newblock \bibinfo{journal}{\emph{IEEE Transactions on Knowledge \& Data
  Engineering}} \bibinfo{number}{6} (\bibinfo{year}{2005}),
  \bibinfo{pages}{734--749}.
\newblock


\bibitem[\protect\citeauthoryear{Adomavicius and Tuzhilin}{Adomavicius and
  Tuzhilin}{2011}]%
        {adomavicius2011context}
\bibfield{author}{\bibinfo{person}{Gediminas Adomavicius} {and}
  \bibinfo{person}{Alexander Tuzhilin}.} \bibinfo{year}{2011}\natexlab{}.
\newblock \showarticletitle{Context-aware recommender systems}.
\newblock In \bibinfo{booktitle}{\emph{Recommender systems handbook}}.
  \bibinfo{publisher}{Springer}, \bibinfo{pages}{217--253}.
\newblock


\bibitem[\protect\citeauthoryear{Arazy, Kumar, and Shapira}{Arazy
  et~al\mbox{.}}{2010}]%
        {arazy2010theory}
\bibfield{author}{\bibinfo{person}{Ofer Arazy}, \bibinfo{person}{Nanda Kumar},
  {and} \bibinfo{person}{Bracha Shapira}.} \bibinfo{year}{2010}\natexlab{}.
\newblock \showarticletitle{A theory-driven design framework for social
  recommender systems}.
\newblock \bibinfo{journal}{\emph{Journal of the Association for Information
  Systems}} \bibinfo{volume}{11}, \bibinfo{number}{9} (\bibinfo{year}{2010}),
  \bibinfo{pages}{455}.
\newblock


\bibitem[\protect\citeauthoryear{Azaria, Kraus, Goldman, and Tsimhoni}{Azaria
  et~al\mbox{.}}{2014}]%
        {azaria2014advice}
\bibfield{author}{\bibinfo{person}{Amos Azaria}, \bibinfo{person}{Sarit Kraus},
  \bibinfo{person}{Claudia~V Goldman}, {and} \bibinfo{person}{Omer Tsimhoni}.}
  \bibinfo{year}{2014}\natexlab{}.
\newblock \showarticletitle{Advice provision for energy saving in automobile
  climate control systems}. In \bibinfo{booktitle}{\emph{Twenty-Sixth IAAI
  Conference}}.
\newblock


\bibitem[\protect\citeauthoryear{Bansback, Brazier, Tsuchiya, and
  Anis}{Bansback et~al\mbox{.}}{2012}]%
        {bansback2012using}
\bibfield{author}{\bibinfo{person}{Nick Bansback}, \bibinfo{person}{John
  Brazier}, \bibinfo{person}{Aki Tsuchiya}, {and} \bibinfo{person}{Aslam
  Anis}.} \bibinfo{year}{2012}\natexlab{}.
\newblock \showarticletitle{Using a discrete choice experiment to estimate
  health state utility values}.
\newblock \bibinfo{journal}{\emph{Journal of health economics}}
  \bibinfo{volume}{31}, \bibinfo{number}{1} (\bibinfo{year}{2012}),
  \bibinfo{pages}{306--318}.
\newblock


\bibitem[\protect\citeauthoryear{Barberis}{Barberis}{2013}]%
        {barberis2013thirty}
\bibfield{author}{\bibinfo{person}{Nicholas~C Barberis}.}
  \bibinfo{year}{2013}\natexlab{}.
\newblock \showarticletitle{Thirty years of prospect theory in economics: A
  review and assessment}.
\newblock \bibinfo{journal}{\emph{Journal of Economic Perspectives}}
  \bibinfo{volume}{27}, \bibinfo{number}{1} (\bibinfo{year}{2013}),
  \bibinfo{pages}{173--96}.
\newblock


\bibitem[\protect\citeauthoryear{Bardzell and Bardzell}{Bardzell and
  Bardzell}{2011}]%
        {bardzell2011towards}
\bibfield{author}{\bibinfo{person}{Shaowen Bardzell} {and}
  \bibinfo{person}{Jeffrey Bardzell}.} \bibinfo{year}{2011}\natexlab{}.
\newblock \showarticletitle{Towards a feminist HCI methodology: social science,
  feminism, and HCI}. In \bibinfo{booktitle}{\emph{Proceedings of the SIGCHI
  Conference on Human Factors in Computing Systems}}. ACM,
  \bibinfo{pages}{675--684}.
\newblock


\bibitem[\protect\citeauthoryear{Benartzi and Thaler}{Benartzi and
  Thaler}{2007}]%
        {benartzi2007heuristics}
\bibfield{author}{\bibinfo{person}{Shlomo Benartzi} {and}
  \bibinfo{person}{Richard Thaler}.} \bibinfo{year}{2007}\natexlab{}.
\newblock \showarticletitle{Heuristics and biases in retirement savings
  behavior}.
\newblock \bibinfo{journal}{\emph{Journal of Economic perspectives}}
  \bibinfo{volume}{21}, \bibinfo{number}{3} (\bibinfo{year}{2007}),
  \bibinfo{pages}{81--104}.
\newblock


\bibitem[\protect\citeauthoryear{Benartzi and Thaler}{Benartzi and
  Thaler}{1999}]%
        {benartzi1999risk}
\bibfield{author}{\bibinfo{person}{Shlomo Benartzi} {and}
  \bibinfo{person}{Richard~H Thaler}.} \bibinfo{year}{1999}\natexlab{}.
\newblock \showarticletitle{Risk aversion or myopia? Choices in repeated
  gambles and retirement investments}.
\newblock \bibinfo{journal}{\emph{Management science}} \bibinfo{volume}{45},
  \bibinfo{number}{3} (\bibinfo{year}{1999}), \bibinfo{pages}{364--381}.
\newblock


\bibitem[\protect\citeauthoryear{Berdichevsky and Neuenschwander}{Berdichevsky
  and Neuenschwander}{1999}]%
        {berdichevsky1999toward}
\bibfield{author}{\bibinfo{person}{Daniel Berdichevsky} {and}
  \bibinfo{person}{Erik Neuenschwander}.} \bibinfo{year}{1999}\natexlab{}.
\newblock \showarticletitle{Toward an ethics of persuasive technology}.
\newblock \bibinfo{journal}{\emph{Commun. ACM}} \bibinfo{volume}{42},
  \bibinfo{number}{5} (\bibinfo{year}{1999}), \bibinfo{pages}{51--58}.
\newblock


\bibitem[\protect\citeauthoryear{Blevis}{Blevis}{2007}]%
        {blevis2007sustainable}
\bibfield{author}{\bibinfo{person}{Eli Blevis}.}
  \bibinfo{year}{2007}\natexlab{}.
\newblock \showarticletitle{Sustainable interaction design: invention \&
  disposal, renewal \& reuse}. In \bibinfo{booktitle}{\emph{Proceedings of the
  SIGCHI conference on Human factors in computing systems}}. ACM,
  \bibinfo{pages}{503--512}.
\newblock


\bibitem[\protect\citeauthoryear{Block and Harper}{Block and Harper}{1991}]%
        {block1991overconfidence}
\bibfield{author}{\bibinfo{person}{Richard~A Block} {and}
  \bibinfo{person}{David~R Harper}.} \bibinfo{year}{1991}\natexlab{}.
\newblock \showarticletitle{Overconfidence in estimation: Testing the
  anchoring-and-adjustment hypothesis}.
\newblock \bibinfo{journal}{\emph{Organizational behavior and human decision
  processes}} \bibinfo{volume}{49}, \bibinfo{number}{2} (\bibinfo{year}{1991}),
  \bibinfo{pages}{188--207}.
\newblock


\bibitem[\protect\citeauthoryear{Bonaccio and Dalal}{Bonaccio and
  Dalal}{2006}]%
        {bonaccio2006advice}
\bibfield{author}{\bibinfo{person}{Silvia Bonaccio} {and}
  \bibinfo{person}{Reeshad~S Dalal}.} \bibinfo{year}{2006}\natexlab{}.
\newblock \showarticletitle{Advice taking and decision-making: An integrative
  literature review, and implications for the organizational sciences}.
\newblock \bibinfo{journal}{\emph{Organizational behavior and human decision
  processes}} \bibinfo{volume}{101}, \bibinfo{number}{2}
  (\bibinfo{year}{2006}), \bibinfo{pages}{127--151}.
\newblock


\bibitem[\protect\citeauthoryear{Bonnefon, Shariff, and Rahwan}{Bonnefon
  et~al\mbox{.}}{2016}]%
        {bonnefon2016social}
\bibfield{author}{\bibinfo{person}{Jean-Fran{\c{c}}ois Bonnefon},
  \bibinfo{person}{Azim Shariff}, {and} \bibinfo{person}{Iyad Rahwan}.}
  \bibinfo{year}{2016}\natexlab{}.
\newblock \showarticletitle{The social dilemma of autonomous vehicles}.
\newblock \bibinfo{journal}{\emph{Science}} \bibinfo{volume}{352},
  \bibinfo{number}{6293} (\bibinfo{year}{2016}), \bibinfo{pages}{1573--1576}.
\newblock


\bibitem[\protect\citeauthoryear{Braun and Clarke}{Braun and Clarke}{2006}]%
        {braun2006using}
\bibfield{author}{\bibinfo{person}{Virginia Braun} {and}
  \bibinfo{person}{Victoria Clarke}.} \bibinfo{year}{2006}\natexlab{}.
\newblock \showarticletitle{Using thematic analysis in psychology}.
\newblock \bibinfo{journal}{\emph{Qualitative research in psychology}}
  \bibinfo{volume}{3}, \bibinfo{number}{2} (\bibinfo{year}{2006}),
  \bibinfo{pages}{77--101}.
\newblock


\bibitem[\protect\citeauthoryear{Brynjolfsson, Collis, and Eggers}{Brynjolfsson
  et~al\mbox{.}}{2019}]%
        {brynjolfsson2019using}
\bibfield{author}{\bibinfo{person}{Erik Brynjolfsson}, \bibinfo{person}{Avinash
  Collis}, {and} \bibinfo{person}{Felix Eggers}.}
  \bibinfo{year}{2019}\natexlab{}.
\newblock \showarticletitle{Using massive online choice experiments to measure
  changes in well-being}.
\newblock \bibinfo{journal}{\emph{Proceedings of the National Academy of
  Sciences}} \bibinfo{volume}{116}, \bibinfo{number}{15}
  (\bibinfo{year}{2019}), \bibinfo{pages}{7250--7255}.
\newblock


\bibitem[\protect\citeauthoryear{Buchanan}{Buchanan}{1985}]%
        {buchanan1985declaration}
\bibfield{author}{\bibinfo{person}{Richard Buchanan}.}
  \bibinfo{year}{1985}\natexlab{}.
\newblock \showarticletitle{Declaration by design: Rhetoric, argument, and
  demonstration in design practice}.
\newblock \bibinfo{journal}{\emph{Design issues}} (\bibinfo{year}{1985}),
  \bibinfo{pages}{4--22}.
\newblock


\bibitem[\protect\citeauthoryear{Buder and Schwind}{Buder and Schwind}{2012}]%
        {buder2012learning}
\bibfield{author}{\bibinfo{person}{J{\"u}rgen Buder} {and}
  \bibinfo{person}{Christina Schwind}.} \bibinfo{year}{2012}\natexlab{}.
\newblock \showarticletitle{Learning with personalized recommender systems: A
  psychological view}.
\newblock \bibinfo{journal}{\emph{Computers in Human Behavior}}
  \bibinfo{volume}{28}, \bibinfo{number}{1} (\bibinfo{year}{2012}),
  \bibinfo{pages}{207--216}.
\newblock


\bibitem[\protect\citeauthoryear{Bursztyn, Ederer, Ferman, and
  Yuchtman}{Bursztyn et~al\mbox{.}}{2014}]%
        {bursztyn2014understanding}
\bibfield{author}{\bibinfo{person}{Leonardo Bursztyn}, \bibinfo{person}{Florian
  Ederer}, \bibinfo{person}{Bruno Ferman}, {and} \bibinfo{person}{Noam
  Yuchtman}.} \bibinfo{year}{2014}\natexlab{}.
\newblock \showarticletitle{Understanding mechanisms underlying peer effects:
  Evidence from a field experiment on financial decisions}.
\newblock \bibinfo{journal}{\emph{Econometrica}} \bibinfo{volume}{82},
  \bibinfo{number}{4} (\bibinfo{year}{2014}), \bibinfo{pages}{1273--1301}.
\newblock


\bibitem[\protect\citeauthoryear{Caraban, Karapanos, Gon{\c{c}}alves, and
  Campos}{Caraban et~al\mbox{.}}{2019}]%
        {caraban201923}
\bibfield{author}{\bibinfo{person}{Ana Caraban}, \bibinfo{person}{Evangelos
  Karapanos}, \bibinfo{person}{Daniel Gon{\c{c}}alves}, {and}
  \bibinfo{person}{Pedro Campos}.} \bibinfo{year}{2019}\natexlab{}.
\newblock \showarticletitle{23 Ways to Nudge: A Review of Technology-Mediated
  Nudging in Human-Computer Interaction}. In
  \bibinfo{booktitle}{\emph{Proceedings of the 2019 CHI Conference on Human
  Factors in Computing Systems}}. ACM, \bibinfo{pages}{503}.
\newblock


\bibitem[\protect\citeauthoryear{Cartwright, Dove, M{\'e}ndez~M{\'e}ndez,
  Bello, and Nov}{Cartwright et~al\mbox{.}}{2019}]%
        {cartwright2019crowdsourcing}
\bibfield{author}{\bibinfo{person}{Mark Cartwright}, \bibinfo{person}{Graham
  Dove}, \bibinfo{person}{Ana~Elisa M{\'e}ndez~M{\'e}ndez},
  \bibinfo{person}{Juan~P Bello}, {and} \bibinfo{person}{Oded Nov}.}
  \bibinfo{year}{2019}\natexlab{}.
\newblock \showarticletitle{Crowdsourcing multi-label audio annotation tasks
  with citizen scientists}. In \bibinfo{booktitle}{\emph{Proceedings of the
  2019 CHI Conference on Human Factors in Computing Systems}}. ACM,
  \bibinfo{pages}{292}.
\newblock


\bibitem[\protect\citeauthoryear{Chang, Harper, and Terveen}{Chang
  et~al\mbox{.}}{2015}]%
        {chang2015using}
\bibfield{author}{\bibinfo{person}{Shuo Chang}, \bibinfo{person}{F~Maxwell
  Harper}, {and} \bibinfo{person}{Loren Terveen}.}
  \bibinfo{year}{2015}\natexlab{}.
\newblock \showarticletitle{Using groups of items to bootstrap new users in
  recommender systems}. In \bibinfo{booktitle}{\emph{CSCW 2015: Proceedings of
  the 18th ACM Conference on Computer Supported Cooperative Work \& Social
  Computing}}. \bibinfo{pages}{1258--1269}.
\newblock


\bibitem[\protect\citeauthoryear{Chivukula, Gray, and Brier}{Chivukula
  et~al\mbox{.}}{2019}]%
        {chivukula2019analyzing}
\bibfield{author}{\bibinfo{person}{Shruthi~Sai Chivukula},
  \bibinfo{person}{Colin~M Gray}, {and} \bibinfo{person}{Jason~A Brier}.}
  \bibinfo{year}{2019}\natexlab{}.
\newblock \showarticletitle{Analyzing Value Discovery in Design Decisions
  Through Ethicography}. In \bibinfo{booktitle}{\emph{Proceedings of the 2019
  CHI Conference on Human Factors in Computing Systems}}. ACM,
  \bibinfo{pages}{77}.
\newblock


\bibitem[\protect\citeauthoryear{Cialdini and Cialdini}{Cialdini and
  Cialdini}{1993}]%
        {cialdini1993influence}
\bibfield{author}{\bibinfo{person}{Robert~B Cialdini} {and}
  \bibinfo{person}{Robert~B Cialdini}.} \bibinfo{year}{1993}\natexlab{}.
\newblock \showarticletitle{Influence: The psychology of persuasion}.
\newblock  (\bibinfo{year}{1993}).
\newblock


\bibitem[\protect\citeauthoryear{Clement and Krueger}{Clement and
  Krueger}{2000}]%
        {clement2000primacy}
\bibfield{author}{\bibinfo{person}{Russell~W Clement} {and}
  \bibinfo{person}{Joachim Krueger}.} \bibinfo{year}{2000}\natexlab{}.
\newblock \showarticletitle{The primacy of self-referent information in
  perceptions of social consensus}.
\newblock \bibinfo{journal}{\emph{British Journal of Social Psychology}}
  \bibinfo{volume}{39}, \bibinfo{number}{2} (\bibinfo{year}{2000}),
  \bibinfo{pages}{279--299}.
\newblock


\bibitem[\protect\citeauthoryear{Daskalova, Lee, Huang, Ni, and
  Lundin}{Daskalova et~al\mbox{.}}{2018}]%
        {daskalova2018investigating}
\bibfield{author}{\bibinfo{person}{Nediyana Daskalova},
  \bibinfo{person}{Bongshin Lee}, \bibinfo{person}{Jeff Huang},
  \bibinfo{person}{Chester Ni}, {and} \bibinfo{person}{Jessica Lundin}.}
  \bibinfo{year}{2018}\natexlab{}.
\newblock \showarticletitle{Investigating the effectiveness of cohort-based
  sleep recommendations}.
\newblock \bibinfo{journal}{\emph{Proceedings of the ACM on Interactive,
  Mobile, Wearable and Ubiquitous Technologies}} \bibinfo{volume}{2},
  \bibinfo{number}{3} (\bibinfo{year}{2018}), \bibinfo{pages}{1--19}.
\newblock


\bibitem[\protect\citeauthoryear{DellaVigna}{DellaVigna}{2009}]%
        {dellavigna2009psychology}
\bibfield{author}{\bibinfo{person}{Stefano DellaVigna}.}
  \bibinfo{year}{2009}\natexlab{}.
\newblock \showarticletitle{Psychology and economics: Evidence from the field}.
\newblock \bibinfo{journal}{\emph{Journal of Economic literature}}
  \bibinfo{volume}{47}, \bibinfo{number}{2} (\bibinfo{year}{2009}),
  \bibinfo{pages}{315--72}.
\newblock


\bibitem[\protect\citeauthoryear{DiSalvo, Sengers, and
  Brynjarsd{\'o}ttir}{DiSalvo et~al\mbox{.}}{2010}]%
        {disalvo2010mapping}
\bibfield{author}{\bibinfo{person}{Carl DiSalvo}, \bibinfo{person}{Phoebe
  Sengers}, {and} \bibinfo{person}{Hr{\"o}nn Brynjarsd{\'o}ttir}.}
  \bibinfo{year}{2010}\natexlab{}.
\newblock \showarticletitle{Mapping the landscape of sustainable HCI}. In
  \bibinfo{booktitle}{\emph{Proceedings of the SIGCHI conference on human
  factors in computing systems}}. ACM, \bibinfo{pages}{1975--1984}.
\newblock


\bibitem[\protect\citeauthoryear{Dunning and Hayes}{Dunning and Hayes}{1996}]%
        {dunning1996evidence}
\bibfield{author}{\bibinfo{person}{David Dunning} {and}
  \bibinfo{person}{Andrew~F Hayes}.} \bibinfo{year}{1996}\natexlab{}.
\newblock \showarticletitle{Evidence for egocentric comparison in social
  judgment.}
\newblock \bibinfo{journal}{\emph{Journal of personality and social
  psychology}} \bibinfo{volume}{71}, \bibinfo{number}{2}
  (\bibinfo{year}{1996}), \bibinfo{pages}{213}.
\newblock


\bibitem[\protect\citeauthoryear{Egesdal, Lai, and Su}{Egesdal
  et~al\mbox{.}}{2015}]%
        {egesdal2015estimating}
\bibfield{author}{\bibinfo{person}{Michael Egesdal}, \bibinfo{person}{Zhenyu
  Lai}, {and} \bibinfo{person}{Che-Lin Su}.} \bibinfo{year}{2015}\natexlab{}.
\newblock \showarticletitle{Estimating dynamic discrete-choice games of
  incomplete information}.
\newblock \bibinfo{journal}{\emph{Quantitative Economics}} \bibinfo{volume}{6},
  \bibinfo{number}{3} (\bibinfo{year}{2015}), \bibinfo{pages}{567--597}.
\newblock


\bibitem[\protect\citeauthoryear{Eysenbach, Powell, Englesakis, Rizo, and
  Stern}{Eysenbach et~al\mbox{.}}{2004}]%
        {eysenbach2004health}
\bibfield{author}{\bibinfo{person}{Gunther Eysenbach}, \bibinfo{person}{John
  Powell}, \bibinfo{person}{Marina Englesakis}, \bibinfo{person}{Carlos Rizo},
  {and} \bibinfo{person}{Anita Stern}.} \bibinfo{year}{2004}\natexlab{}.
\newblock \showarticletitle{Health related virtual communities and electronic
  support groups: systematic review of the effects of online peer to peer
  interactions}.
\newblock \bibinfo{journal}{\emph{Bmj}} \bibinfo{volume}{328},
  \bibinfo{number}{7449} (\bibinfo{year}{2004}), \bibinfo{pages}{1166}.
\newblock


\bibitem[\protect\citeauthoryear{Fleischer, Mead, and Huang}{Fleischer
  et~al\mbox{.}}{2015}]%
        {fleischer2015inattentive}
\bibfield{author}{\bibinfo{person}{Avi Fleischer}, \bibinfo{person}{Alan~D
  Mead}, {and} \bibinfo{person}{Jialin Huang}.}
  \bibinfo{year}{2015}\natexlab{}.
\newblock \showarticletitle{Inattentive responding in MTurk and other online
  samples}.
\newblock \bibinfo{journal}{\emph{Industrial and Organizational Psychology}}
  \bibinfo{volume}{8}, \bibinfo{number}{2} (\bibinfo{year}{2015}),
  \bibinfo{pages}{196--202}.
\newblock


\bibitem[\protect\citeauthoryear{Friedman and Kahn~Jr}{Friedman and
  Kahn~Jr}{2003}]%
        {friedman2003human}
\bibfield{author}{\bibinfo{person}{Batya Friedman} {and}
  \bibinfo{person}{Peter~H Kahn~Jr}.} \bibinfo{year}{2003}\natexlab{}.
\newblock \showarticletitle{Human values, ethics, and design}.
\newblock \bibinfo{journal}{\emph{The human-computer interaction handbook}}
  (\bibinfo{year}{2003}), \bibinfo{pages}{1177--1201}.
\newblock


\bibitem[\protect\citeauthoryear{Galloway}{Galloway}{[n. d.]}]%
        {gallwayrun}
\bibfield{author}{\bibinfo{person}{Jeff Galloway}.} \bibinfo{year}{[n.
  d.]}\natexlab{}.
\newblock \bibinfo{title}{Training: Run, Walk, Run}.
\newblock
\newblock
\urldef\tempurl%
\url{http://www.jeffgalloway.com/training/run-walk/}
\showURL{%
Retrieved August 28, 2019 from \tempurl}


\bibitem[\protect\citeauthoryear{Goodwin and Fildes}{Goodwin and
  Fildes}{1999}]%
        {goodwin1999judgmental}
\bibfield{author}{\bibinfo{person}{Paul Goodwin} {and} \bibinfo{person}{Robert
  Fildes}.} \bibinfo{year}{1999}\natexlab{}.
\newblock \showarticletitle{Judgmental forecasts of time series affected by
  special events: Does providing a statistical forecast improve accuracy?}
\newblock \bibinfo{journal}{\emph{Journal of Behavioral Decision Making}}
  \bibinfo{volume}{12}, \bibinfo{number}{1} (\bibinfo{year}{1999}),
  \bibinfo{pages}{37--53}.
\newblock


\bibitem[\protect\citeauthoryear{Gunaratne and Nov}{Gunaratne and Nov}{2015a}]%
        {gunaratne2015influencing}
\bibfield{author}{\bibinfo{person}{Junius Gunaratne} {and}
  \bibinfo{person}{Oded Nov}.} \bibinfo{year}{2015}\natexlab{a}.
\newblock \showarticletitle{Influencing retirement saving behavior with expert
  advice and social comparison as persuasive techniques}. In
  \bibinfo{booktitle}{\emph{International Conference on Persuasive
  Technology}}. Springer, \bibinfo{pages}{205--216}.
\newblock


\bibitem[\protect\citeauthoryear{Gunaratne and Nov}{Gunaratne and Nov}{2015b}]%
        {gunaratne2015informing}
\bibfield{author}{\bibinfo{person}{Junius Gunaratne} {and}
  \bibinfo{person}{Oded Nov}.} \bibinfo{year}{2015}\natexlab{b}.
\newblock \showarticletitle{Informing and improving retirement saving
  performance using behavioral economics theory-driven user interfaces}. In
  \bibinfo{booktitle}{\emph{Proceedings of the 33rd annual ACM conference on
  human factors in computing systems}}. ACM, \bibinfo{pages}{917--920}.
\newblock


\bibitem[\protect\citeauthoryear{Gunaratne, Zalmanson, and Nov}{Gunaratne
  et~al\mbox{.}}{2018}]%
        {gunaratne2018persuasive}
\bibfield{author}{\bibinfo{person}{Junius Gunaratne}, \bibinfo{person}{Lior
  Zalmanson}, {and} \bibinfo{person}{Oded Nov}.}
  \bibinfo{year}{2018}\natexlab{}.
\newblock \showarticletitle{The Persuasive Power of Algorithmic and
  Crowdsourced Advice}.
\newblock \bibinfo{journal}{\emph{Journal of Management Information Systems}}
  \bibinfo{volume}{35}, \bibinfo{number}{4} (\bibinfo{year}{2018}),
  \bibinfo{pages}{1092--1120}.
\newblock


\bibitem[\protect\citeauthoryear{Hacker and O'Leary}{Hacker and
  O'Leary}{2012}]%
        {hacker2012shared}
\bibfield{author}{\bibinfo{person}{Jacob Hacker} {and} \bibinfo{person}{Ann
  O'Leary}.} \bibinfo{year}{2012}\natexlab{}.
\newblock \bibinfo{booktitle}{\emph{Shared responsibility, shared risk:
  government, markets and social policy in the twenty-first century}}.
\newblock \bibinfo{publisher}{OUP USA}.
\newblock


\bibitem[\protect\citeauthoryear{Harman, O'Donovan, Abdelzaher, and
  Gonzalez}{Harman et~al\mbox{.}}{2014}]%
        {harman2014dynamics}
\bibfield{author}{\bibinfo{person}{Jason~L Harman}, \bibinfo{person}{John
  O'Donovan}, \bibinfo{person}{Tarek Abdelzaher}, {and}
  \bibinfo{person}{Cleotilde Gonzalez}.} \bibinfo{year}{2014}\natexlab{}.
\newblock \showarticletitle{Dynamics of human trust in recommender systems}. In
  \bibinfo{booktitle}{\emph{Proceedings of the 8th ACM Conference on
  Recommender systems}}. \bibinfo{pages}{305--308}.
\newblock


\bibitem[\protect\citeauthoryear{Hartmann, MacDougall, Brandt, and
  Klemmer}{Hartmann et~al\mbox{.}}{2010}]%
        {hartmann2010would}
\bibfield{author}{\bibinfo{person}{Bj{\"o}rn Hartmann}, \bibinfo{person}{Daniel
  MacDougall}, \bibinfo{person}{Joel Brandt}, {and} \bibinfo{person}{Scott~R
  Klemmer}.} \bibinfo{year}{2010}\natexlab{}.
\newblock \showarticletitle{What would other programmers do: suggesting
  solutions to error messages}. In \bibinfo{booktitle}{\emph{Proceedings of the
  SIGCHI Conference on Human Factors in Computing Systems}}. ACM,
  \bibinfo{pages}{1019--1028}.
\newblock


\bibitem[\protect\citeauthoryear{Hauser and Schwarz}{Hauser and
  Schwarz}{2016}]%
        {hauser2016attentive}
\bibfield{author}{\bibinfo{person}{David~J Hauser} {and}
  \bibinfo{person}{Norbert Schwarz}.} \bibinfo{year}{2016}\natexlab{}.
\newblock \showarticletitle{Attentive Turkers: MTurk participants perform
  better on online attention checks than do subject pool participants}.
\newblock \bibinfo{journal}{\emph{Behavior research methods}}
  \bibinfo{volume}{48}, \bibinfo{number}{1} (\bibinfo{year}{2016}),
  \bibinfo{pages}{400--407}.
\newblock


\bibitem[\protect\citeauthoryear{Herlocker, Konstan, and Riedl}{Herlocker
  et~al\mbox{.}}{2000}]%
        {herlocker2000explaining}
\bibfield{author}{\bibinfo{person}{Jonathan~L Herlocker},
  \bibinfo{person}{Joseph~A Konstan}, {and} \bibinfo{person}{John Riedl}.}
  \bibinfo{year}{2000}\natexlab{}.
\newblock \showarticletitle{Explaining collaborative filtering
  recommendations}. In \bibinfo{booktitle}{\emph{Proceedings of the 2000 ACM
  conference on Computer supported cooperative work}}. ACM,
  \bibinfo{pages}{241--250}.
\newblock


\bibitem[\protect\citeauthoryear{Horne, Jaccard, and Tiedemann}{Horne
  et~al\mbox{.}}{2005}]%
        {horne2005improving}
\bibfield{author}{\bibinfo{person}{Matt Horne}, \bibinfo{person}{Mark Jaccard},
  {and} \bibinfo{person}{Ken Tiedemann}.} \bibinfo{year}{2005}\natexlab{}.
\newblock \showarticletitle{Improving behavioral realism in hybrid
  energy-economy models using discrete choice studies of personal
  transportation decisions}.
\newblock \bibinfo{journal}{\emph{Energy Economics}} \bibinfo{volume}{27},
  \bibinfo{number}{1} (\bibinfo{year}{2005}), \bibinfo{pages}{59--77}.
\newblock


\bibitem[\protect\citeauthoryear{Jannach, Resnick, Tuzhilin, and
  Zanker}{Jannach et~al\mbox{.}}{2016}]%
        {jannach2016recommender}
\bibfield{author}{\bibinfo{person}{Dietmar Jannach}, \bibinfo{person}{Paul
  Resnick}, \bibinfo{person}{Alexander Tuzhilin}, {and} \bibinfo{person}{Markus
  Zanker}.} \bibinfo{year}{2016}\natexlab{}.
\newblock \showarticletitle{Recommender systems—beyond matrix completion}.
\newblock \bibinfo{journal}{\emph{Commun. ACM}} \bibinfo{volume}{59},
  \bibinfo{number}{11} (\bibinfo{year}{2016}), \bibinfo{pages}{94--102}.
\newblock


\bibitem[\protect\citeauthoryear{Jonas, Schulz-Hardt, Frey, and Thelen}{Jonas
  et~al\mbox{.}}{2001}]%
        {jonas2001confirmation}
\bibfield{author}{\bibinfo{person}{Eva Jonas}, \bibinfo{person}{Stefan
  Schulz-Hardt}, \bibinfo{person}{Dieter Frey}, {and} \bibinfo{person}{Norman
  Thelen}.} \bibinfo{year}{2001}\natexlab{}.
\newblock \showarticletitle{Confirmation bias in sequential information search
  after preliminary decisions: an expansion of dissonance theoretical research
  on selective exposure to information.}
\newblock \bibinfo{journal}{\emph{Journal of personality and social
  psychology}} \bibinfo{volume}{80}, \bibinfo{number}{4}
  (\bibinfo{year}{2001}), \bibinfo{pages}{557}.
\newblock


\bibitem[\protect\citeauthoryear{Jowett and O'donnell}{Jowett and
  O'donnell}{2018}]%
        {jowett2018propaganda}
\bibfield{author}{\bibinfo{person}{Garth~S Jowett} {and}
  \bibinfo{person}{Victoria O'donnell}.} \bibinfo{year}{2018}\natexlab{}.
\newblock \bibinfo{booktitle}{\emph{Propaganda \& persuasion}}.
\newblock \bibinfo{publisher}{Sage Publications}.
\newblock


\bibitem[\protect\citeauthoryear{Kahneman and Thaler}{Kahneman and
  Thaler}{2006}]%
        {kahneman2006anomalies}
\bibfield{author}{\bibinfo{person}{Daniel Kahneman} {and}
  \bibinfo{person}{Richard~H Thaler}.} \bibinfo{year}{2006}\natexlab{}.
\newblock \showarticletitle{Anomalies: Utility maximization and experienced
  utility}.
\newblock \bibinfo{journal}{\emph{Journal of Economic Perspectives}}
  \bibinfo{volume}{20}, \bibinfo{number}{1} (\bibinfo{year}{2006}),
  \bibinfo{pages}{221--234}.
\newblock


\bibitem[\protect\citeauthoryear{Kahneman and Tversky}{Kahneman and
  Tversky}{1979}]%
        {kahneman979prospect}
\bibfield{author}{\bibinfo{person}{Daniel Kahneman} {and} \bibinfo{person}{Amos
  Tversky}.} \bibinfo{year}{1979}\natexlab{}.
\newblock \showarticletitle{Prospect theory: An analysis of decision under
  risk}.
\newblock \bibinfo{journal}{\emph{Econometrica}} \bibinfo{volume}{47},
  \bibinfo{number}{2} (\bibinfo{year}{1979}), \bibinfo{pages}{363--391}.
\newblock


\bibitem[\protect\citeauthoryear{Kapoor, Kumar, Terveen, Konstan, and
  Schrater}{Kapoor et~al\mbox{.}}{2015}]%
        {kapoor2015like}
\bibfield{author}{\bibinfo{person}{Komal Kapoor}, \bibinfo{person}{Vikas
  Kumar}, \bibinfo{person}{Loren Terveen}, \bibinfo{person}{Joseph~A Konstan},
  {and} \bibinfo{person}{Paul Schrater}.} \bibinfo{year}{2015}\natexlab{}.
\newblock \showarticletitle{" I like to explore sometimes" Adapting to Dynamic
  User Novelty Preferences}. In \bibinfo{booktitle}{\emph{Proceedings of the
  9th ACM Conference on Recommender Systems}}. \bibinfo{pages}{19--26}.
\newblock


\bibitem[\protect\citeauthoryear{Kehr, Hassenzahl, Laschke, and
  Diefenbach}{Kehr et~al\mbox{.}}{2012}]%
        {kehr2012transformational}
\bibfield{author}{\bibinfo{person}{Flavius Kehr}, \bibinfo{person}{Marc
  Hassenzahl}, \bibinfo{person}{Matthias Laschke}, {and} \bibinfo{person}{Sarah
  Diefenbach}.} \bibinfo{year}{2012}\natexlab{}.
\newblock \showarticletitle{A transformational product to improve self-control
  strength: the chocolate machine}. In \bibinfo{booktitle}{\emph{Proceedings of
  the SIGCHI Conference on Human Factors in Computing Systems}}. ACM,
  \bibinfo{pages}{689--694}.
\newblock


\bibitem[\protect\citeauthoryear{Koivunen, Olsson, Olshannikova, and
  Lindberg}{Koivunen et~al\mbox{.}}{2019}]%
        {koivunen2019understanding}
\bibfield{author}{\bibinfo{person}{Sami Koivunen}, \bibinfo{person}{Thomas
  Olsson}, \bibinfo{person}{Ekaterina Olshannikova}, {and} \bibinfo{person}{Aki
  Lindberg}.} \bibinfo{year}{2019}\natexlab{}.
\newblock \showarticletitle{Understanding Decision-Making in Recruitment:
  Opportunities and Challenges for Information Technology}.
\newblock \bibinfo{journal}{\emph{Proceedings of the ACM on Human-Computer
  Interaction}} \bibinfo{volume}{3}, \bibinfo{number}{GROUP}
  (\bibinfo{year}{2019}), \bibinfo{pages}{1--22}.
\newblock


\bibitem[\protect\citeauthoryear{Krueger and Stanke}{Krueger and
  Stanke}{2001}]%
        {krueger2001role}
\bibfield{author}{\bibinfo{person}{Joachim Krueger} {and}
  \bibinfo{person}{David Stanke}.} \bibinfo{year}{2001}\natexlab{}.
\newblock \showarticletitle{The role of self-referent and other-referent
  knowledge in perceptions of group characteristics}.
\newblock \bibinfo{journal}{\emph{Personality and Social Psychology Bulletin}}
  \bibinfo{volume}{27}, \bibinfo{number}{7} (\bibinfo{year}{2001}),
  \bibinfo{pages}{878--888}.
\newblock


\bibitem[\protect\citeauthoryear{Kruger}{Kruger}{1999}]%
        {kruger1999lake}
\bibfield{author}{\bibinfo{person}{Justin Kruger}.}
  \bibinfo{year}{1999}\natexlab{}.
\newblock \showarticletitle{Lake Wobegon be gone! The below-average effect and
  the egocentric nature of comparative ability judgments.}
\newblock \bibinfo{journal}{\emph{Journal of personality and social
  psychology}} \bibinfo{volume}{77}, \bibinfo{number}{2}
  (\bibinfo{year}{1999}), \bibinfo{pages}{221}.
\newblock


\bibitem[\protect\citeauthoryear{Lacke}{Lacke}{2017}]%
        {lacke2017run}
\bibfield{author}{\bibinfo{person}{Susan Lacke}.}
  \bibinfo{year}{2017}\natexlab{}.
\newblock \bibinfo{title}{3 Run/Walk Interval Workouts To Incorporate Into Your
  Training}.
\newblock
\newblock
\urldef\tempurl%
\url{https://www.womensrunning.com/2017/09/training/run-walk-interval-workouts_81066}
\showURL{%
Retrieved August 28, 2019 from \tempurl}


\bibitem[\protect\citeauthoryear{Landers and Behrend}{Landers and
  Behrend}{2015}]%
        {landers2015inconvenient}
\bibfield{author}{\bibinfo{person}{Richard~N Landers} {and}
  \bibinfo{person}{Tara~S Behrend}.} \bibinfo{year}{2015}\natexlab{}.
\newblock \showarticletitle{An inconvenient truth: Arbitrary distinctions
  between organizational, Mechanical Turk, and other convenience samples}.
\newblock \bibinfo{journal}{\emph{Industrial and Organizational Psychology}}
  \bibinfo{volume}{8}, \bibinfo{number}{2} (\bibinfo{year}{2015}),
  \bibinfo{pages}{142--164}.
\newblock


\bibitem[\protect\citeauthoryear{Lee, Kiesler, and Forlizzi}{Lee
  et~al\mbox{.}}{2011}]%
        {lee2011mining}
\bibfield{author}{\bibinfo{person}{Min~Kyung Lee}, \bibinfo{person}{Sara
  Kiesler}, {and} \bibinfo{person}{Jodi Forlizzi}.}
  \bibinfo{year}{2011}\natexlab{}.
\newblock \showarticletitle{Mining behavioral economics to design persuasive
  technology for healthy choices}. In \bibinfo{booktitle}{\emph{Proceedings of
  the SIGCHI Conference on Human Factors in Computing Systems}}. ACM,
  \bibinfo{pages}{325--334}.
\newblock


\bibitem[\protect\citeauthoryear{Levy and Sarne}{Levy and Sarne}{2016}]%
        {levy2016intelligent}
\bibfield{author}{\bibinfo{person}{Priel Levy} {and} \bibinfo{person}{David
  Sarne}.} \bibinfo{year}{2016}\natexlab{}.
\newblock \showarticletitle{Intelligent advice provisioning for repeated
  interaction}. In \bibinfo{booktitle}{\emph{Thirtieth AAAI Conference on
  Artificial Intelligence}}.
\newblock


\bibitem[\protect\citeauthoryear{Lim and O'Connor}{Lim and O'Connor}{1995}]%
        {lim1995judgemental}
\bibfield{author}{\bibinfo{person}{Joa~Sang Lim} {and} \bibinfo{person}{Marcus
  O'Connor}.} \bibinfo{year}{1995}\natexlab{}.
\newblock \showarticletitle{Judgemental adjustment of initial forecasts: Its
  effectiveness and biases}.
\newblock \bibinfo{journal}{\emph{Journal of Behavioral Decision Making}}
  \bibinfo{volume}{8}, \bibinfo{number}{3} (\bibinfo{year}{1995}),
  \bibinfo{pages}{149--168}.
\newblock


\bibitem[\protect\citeauthoryear{Markus and Kitayama}{Markus and
  Kitayama}{1991}]%
        {markus1991culture}
\bibfield{author}{\bibinfo{person}{Hazel~R Markus} {and}
  \bibinfo{person}{Shinobu Kitayama}.} \bibinfo{year}{1991}\natexlab{}.
\newblock \showarticletitle{Culture and the self: Implications for cognition,
  emotion, and motivation.}
\newblock \bibinfo{journal}{\emph{Psychological review}} \bibinfo{volume}{98},
  \bibinfo{number}{2} (\bibinfo{year}{1991}), \bibinfo{pages}{224}.
\newblock


\bibitem[\protect\citeauthoryear{McNee, Riedl, and Konstan}{McNee
  et~al\mbox{.}}{2006}]%
        {mcnee2006being}
\bibfield{author}{\bibinfo{person}{Sean~M McNee}, \bibinfo{person}{John Riedl},
  {and} \bibinfo{person}{Joseph~A Konstan}.} \bibinfo{year}{2006}\natexlab{}.
\newblock \showarticletitle{Being accurate is not enough: how accuracy metrics
  have hurt recommender systems}. In \bibinfo{booktitle}{\emph{CHI'06 extended
  abstracts on Human factors in computing systems}}. ACM,
  \bibinfo{pages}{1097--1101}.
\newblock


\bibitem[\protect\citeauthoryear{Milano, Taddeo, and Floridi}{Milano
  et~al\mbox{.}}{2020}]%
        {milano2020recommender}
\bibfield{author}{\bibinfo{person}{Silvia Milano},
  \bibinfo{person}{Mariarosaria Taddeo}, {and} \bibinfo{person}{Luciano
  Floridi}.} \bibinfo{year}{2020}\natexlab{}.
\newblock \showarticletitle{Recommender systems and their ethical challenges}.
\newblock \bibinfo{journal}{\emph{AI \& SOCIETY}} (\bibinfo{year}{2020}),
  \bibinfo{pages}{1--11}.
\newblock


\bibitem[\protect\citeauthoryear{Milgram}{Milgram}{1963}]%
        {milgram1963behavioral}
\bibfield{author}{\bibinfo{person}{Stanley Milgram}.}
  \bibinfo{year}{1963}\natexlab{}.
\newblock \showarticletitle{Behavioral study of obedience.}
\newblock \bibinfo{journal}{\emph{The Journal of abnormal and social
  psychology}} \bibinfo{volume}{67}, \bibinfo{number}{4}
  (\bibinfo{year}{1963}), \bibinfo{pages}{371}.
\newblock


\bibitem[\protect\citeauthoryear{Moore, Chen, Turnbull, and Joachims}{Moore
  et~al\mbox{.}}{2013}]%
        {moore2013taste}
\bibfield{author}{\bibinfo{person}{Joshua~L Moore}, \bibinfo{person}{Shuo
  Chen}, \bibinfo{person}{Douglas Turnbull}, {and} \bibinfo{person}{Thorsten
  Joachims}.} \bibinfo{year}{2013}\natexlab{}.
\newblock \showarticletitle{Taste Over Time: The Temporal Dynamics of User
  Preferences.}. In \bibinfo{booktitle}{\emph{ISMIR}}. Citeseer,
  \bibinfo{pages}{401--406}.
\newblock


\bibitem[\protect\citeauthoryear{Nebeker, Torous, and Ellis}{Nebeker
  et~al\mbox{.}}{2019}]%
        {nebeker2019building}
\bibfield{author}{\bibinfo{person}{Camille Nebeker}, \bibinfo{person}{John
  Torous}, {and} \bibinfo{person}{Rebecca J~Bartlett Ellis}.}
  \bibinfo{year}{2019}\natexlab{}.
\newblock \showarticletitle{Building the case for actionable ethics in digital
  health research supported by artificial intelligence}.
\newblock \bibinfo{journal}{\emph{BMC medicine}} \bibinfo{volume}{17},
  \bibinfo{number}{1} (\bibinfo{year}{2019}), \bibinfo{pages}{137}.
\newblock


\bibitem[\protect\citeauthoryear{Nguyen, Dabbish, and Kiesler}{Nguyen
  et~al\mbox{.}}{2015}]%
        {nguyen2015perverse}
\bibfield{author}{\bibinfo{person}{Duyen~T Nguyen}, \bibinfo{person}{Laura~A
  Dabbish}, {and} \bibinfo{person}{Sara Kiesler}.}
  \bibinfo{year}{2015}\natexlab{}.
\newblock \showarticletitle{The perverse effects of social transparency on
  online advice taking}. In \bibinfo{booktitle}{\emph{Proceedings of the 18th
  ACM Conference on Computer Supported Cooperative Work \& Social Computing}}.
  ACM, \bibinfo{pages}{207--217}.
\newblock


\bibitem[\protect\citeauthoryear{Nov and Arazy}{Nov and Arazy}{2013}]%
        {nov2013personality}
\bibfield{author}{\bibinfo{person}{Oded Nov} {and} \bibinfo{person}{Ofer
  Arazy}.} \bibinfo{year}{2013}\natexlab{}.
\newblock \showarticletitle{Personality-targeted design: theory, experimental
  procedure, and preliminary results}. In \bibinfo{booktitle}{\emph{Proceedings
  of the ACM Conference on Computer Supported Cooperative Work.}} ACM,
  \bibinfo{pages}{977--984}.
\newblock


\bibitem[\protect\citeauthoryear{of~Health, Services, and
  of~Agriculture}{of~Health et~al\mbox{.}}{2015}]%
        {usda2015diet}
\bibfield{author}{\bibinfo{person}{U.S.~Department of Health},
  \bibinfo{person}{Human Services}, {and} \bibinfo{person}{U.S.~Department of
  Agriculture}.} \bibinfo{year}{2015}\natexlab{}.
\newblock \bibinfo{title}{Dietary Guidelines for Americans}.
\newblock
\newblock
\urldef\tempurl%
\url{ttps://health.gov/dietaryguidelines/2015/guidelines/}
\showURL{%
Retrieved August 28, 2019 from \tempurl}


\bibitem[\protect\citeauthoryear{{\"O}nkal, Goodwin, Thomson, G{\"o}n{\"u}l,
  and Pollock}{{\"O}nkal et~al\mbox{.}}{2009}]%
        {onkal2009relative}
\bibfield{author}{\bibinfo{person}{Dilek {\"O}nkal}, \bibinfo{person}{Paul
  Goodwin}, \bibinfo{person}{Mary Thomson}, \bibinfo{person}{Sinan
  G{\"o}n{\"u}l}, {and} \bibinfo{person}{Andrew Pollock}.}
  \bibinfo{year}{2009}\natexlab{}.
\newblock \showarticletitle{The relative influence of advice from human experts
  and statistical methods on forecast adjustments}.
\newblock \bibinfo{journal}{\emph{Journal of Behavioral Decision Making}}
  \bibinfo{volume}{22}, \bibinfo{number}{4} (\bibinfo{year}{2009}),
  \bibinfo{pages}{390--409}.
\newblock


\bibitem[\protect\citeauthoryear{Paraschakis}{Paraschakis}{2016}]%
        {paraschakis2016recommender}
\bibfield{author}{\bibinfo{person}{Dimitris Paraschakis}.}
  \bibinfo{year}{2016}\natexlab{}.
\newblock \showarticletitle{Recommender systems from an industrial and ethical
  perspective}. In \bibinfo{booktitle}{\emph{Proceedings of the 10th ACM
  conference on recommender systems}}. \bibinfo{pages}{463--466}.
\newblock


\bibitem[\protect\citeauthoryear{Purpura, Schwanda, Williams, Stubler, and
  Sengers}{Purpura et~al\mbox{.}}{2011}]%
        {purpura2011fit4life}
\bibfield{author}{\bibinfo{person}{Stephen Purpura}, \bibinfo{person}{Victoria
  Schwanda}, \bibinfo{person}{Kaiton Williams}, \bibinfo{person}{William
  Stubler}, {and} \bibinfo{person}{Phoebe Sengers}.}
  \bibinfo{year}{2011}\natexlab{}.
\newblock \showarticletitle{Fit4life: the design of a persuasive technology
  promoting healthy behavior and ideal weight}. In
  \bibinfo{booktitle}{\emph{Proceedings of the SIGCHI conference on human
  factors in computing systems}}. ACM, \bibinfo{pages}{423--432}.
\newblock


\bibitem[\protect\citeauthoryear{Quadrana and Cremonesi}{Quadrana and
  Cremonesi}{2018}]%
        {quadrana2018sequence2}
\bibfield{author}{\bibinfo{person}{Massimo Quadrana} {and}
  \bibinfo{person}{Paolo Cremonesi}.} \bibinfo{year}{2018}\natexlab{}.
\newblock \showarticletitle{Sequence-aware recommendation}. In
  \bibinfo{booktitle}{\emph{Proceedings of the 12th ACM Conference on
  Recommender Systems}}. \bibinfo{pages}{539--540}.
\newblock


\bibitem[\protect\citeauthoryear{Quadrana, Cremonesi, and Jannach}{Quadrana
  et~al\mbox{.}}{2018}]%
        {quadrana2018sequence}
\bibfield{author}{\bibinfo{person}{Massimo Quadrana}, \bibinfo{person}{Paolo
  Cremonesi}, {and} \bibinfo{person}{Dietmar Jannach}.}
  \bibinfo{year}{2018}\natexlab{}.
\newblock \showarticletitle{Sequence-aware recommender systems}.
\newblock \bibinfo{journal}{\emph{ACM Computing Surveys (CSUR)}}
  \bibinfo{volume}{51}, \bibinfo{number}{4} (\bibinfo{year}{2018}),
  \bibinfo{pages}{1--36}.
\newblock


\bibitem[\protect\citeauthoryear{Redstr{\"o}m}{Redstr{\"o}m}{2006}]%
        {redstrom2006persuasive}
\bibfield{author}{\bibinfo{person}{Johan Redstr{\"o}m}.}
  \bibinfo{year}{2006}\natexlab{}.
\newblock \showarticletitle{Persuasive design: Fringes and foundations}. In
  \bibinfo{booktitle}{\emph{International Conference on Persuasive
  Technology}}. Springer, \bibinfo{pages}{112--122}.
\newblock


\bibitem[\protect\citeauthoryear{Samson and Sumi}{Samson and Sumi}{2019}]%
        {samson2019exploring}
\bibfield{author}{\bibinfo{person}{Briane Paul~V Samson} {and}
  \bibinfo{person}{Yasuyuki Sumi}.} \bibinfo{year}{2019}\natexlab{}.
\newblock \showarticletitle{Exploring Factors that Influence Connected Drivers
  to (Not) Use or Follow Recommended Optimal Routes}. In
  \bibinfo{booktitle}{\emph{Proceedings of the 2019 CHI Conference on Human
  Factors in Computing Systems}}. ACM, \bibinfo{pages}{371}.
\newblock


\bibitem[\protect\citeauthoryear{Schwind, Buder, and Hesse}{Schwind
  et~al\mbox{.}}{2011}]%
        {schwind2011will}
\bibfield{author}{\bibinfo{person}{Christina Schwind},
  \bibinfo{person}{J{\"u}rgen Buder}, {and} \bibinfo{person}{Friedrich~W
  Hesse}.} \bibinfo{year}{2011}\natexlab{}.
\newblock \showarticletitle{I will do it, but i don't like it: user reactions
  to preference-inconsistent recommendations}. In
  \bibinfo{booktitle}{\emph{Proceedings of the SIGCHI Conference on Human
  Factors in Computing Systems}}. ACM, \bibinfo{pages}{349--352}.
\newblock


\bibitem[\protect\citeauthoryear{Shefrin and Thaler}{Shefrin and
  Thaler}{1988}]%
        {shefrin1988behavioral}
\bibfield{author}{\bibinfo{person}{Hersh~M Shefrin} {and}
  \bibinfo{person}{Richard~H Thaler}.} \bibinfo{year}{1988}\natexlab{}.
\newblock \showarticletitle{The behavioral life-cycle hypothesis}.
\newblock \bibinfo{journal}{\emph{Economic inquiry}} \bibinfo{volume}{26},
  \bibinfo{number}{4} (\bibinfo{year}{1988}), \bibinfo{pages}{609--643}.
\newblock


\bibitem[\protect\citeauthoryear{Swearingen and Sinha}{Swearingen and
  Sinha}{2001}]%
        {swearingen2001beyond}
\bibfield{author}{\bibinfo{person}{Kirsten Swearingen} {and}
  \bibinfo{person}{Rashmi Sinha}.} \bibinfo{year}{2001}\natexlab{}.
\newblock \showarticletitle{Beyond algorithms: An HCI perspective on
  recommender systems}. In \bibinfo{booktitle}{\emph{ACM SIGIR 2001 workshop on
  recommender systems}}, Vol.~\bibinfo{volume}{13}. Citeseer,
  \bibinfo{pages}{1--11}.
\newblock


\bibitem[\protect\citeauthoryear{Thaler}{Thaler}{2018}]%
        {thaler2018cashews}
\bibfield{author}{\bibinfo{person}{Richard~H Thaler}.}
  \bibinfo{year}{2018}\natexlab{}.
\newblock \showarticletitle{From cashews to nudges: the evolution of behavioral
  economics}.
\newblock \bibinfo{journal}{\emph{American Economic Review}}
  \bibinfo{volume}{108}, \bibinfo{number}{6} (\bibinfo{year}{2018}),
  \bibinfo{pages}{1265--87}.
\newblock


\bibitem[\protect\citeauthoryear{Thaler and Benartzi}{Thaler and
  Benartzi}{2007}]%
        {thaler2007behavioral}
\bibfield{author}{\bibinfo{person}{Richard~H Thaler} {and}
  \bibinfo{person}{Shlomo Benartzi}.} \bibinfo{year}{2007}\natexlab{}.
\newblock \showarticletitle{The behavioral economics of retirement savings
  behavior}.
\newblock  (\bibinfo{year}{2007}).
\newblock


\bibitem[\protect\citeauthoryear{Thomson}{Thomson}{1976}]%
        {thomson1976killing}
\bibfield{author}{\bibinfo{person}{Judith~Jarvis Thomson}.}
  \bibinfo{year}{1976}\natexlab{}.
\newblock \showarticletitle{Killing, letting die, and the trolley problem}.
\newblock \bibinfo{journal}{\emph{The Monist}} \bibinfo{volume}{59},
  \bibinfo{number}{2} (\bibinfo{year}{1976}), \bibinfo{pages}{204--217}.
\newblock


\bibitem[\protect\citeauthoryear{Triandis, McCusker, and Hui}{Triandis
  et~al\mbox{.}}{1990}]%
        {triandis1990multimethod}
\bibfield{author}{\bibinfo{person}{Harry~C Triandis},
  \bibinfo{person}{Christopher McCusker}, {and} \bibinfo{person}{C~Harry Hui}.}
  \bibinfo{year}{1990}\natexlab{}.
\newblock \showarticletitle{Multimethod probes of individualism and
  collectivism.}
\newblock \bibinfo{journal}{\emph{Journal of personality and social
  psychology}} \bibinfo{volume}{59}, \bibinfo{number}{5}
  (\bibinfo{year}{1990}), \bibinfo{pages}{1006}.
\newblock


\bibitem[\protect\citeauthoryear{Tsai and Brusilovsky}{Tsai and
  Brusilovsky}{2017}]%
        {tsai2017providing}
\bibfield{author}{\bibinfo{person}{Chun-Hua Tsai} {and} \bibinfo{person}{Peter
  Brusilovsky}.} \bibinfo{year}{2017}\natexlab{}.
\newblock \showarticletitle{Providing control and transparency in a social
  recommender system for academic conferences}. In
  \bibinfo{booktitle}{\emph{Proceedings of the 25th Conference on User
  Modeling, Adaptation and Personalization}}. \bibinfo{pages}{313--317}.
\newblock


\bibitem[\protect\citeauthoryear{Tversky and Kahneman}{Tversky and
  Kahneman}{1974}]%
        {tversky1974judgment}
\bibfield{author}{\bibinfo{person}{Amos Tversky} {and} \bibinfo{person}{Daniel
  Kahneman}.} \bibinfo{year}{1974}\natexlab{}.
\newblock \showarticletitle{Judgment under uncertainty: Heuristics and biases}.
\newblock \bibinfo{journal}{\emph{science}} \bibinfo{volume}{185},
  \bibinfo{number}{4157} (\bibinfo{year}{1974}), \bibinfo{pages}{1124--1131}.
\newblock


\bibitem[\protect\citeauthoryear{Tyler and Lind}{Tyler and Lind}{1992}]%
        {tyler1992relational}
\bibfield{author}{\bibinfo{person}{Tom~R Tyler} {and} \bibinfo{person}{E~Allan
  Lind}.} \bibinfo{year}{1992}\natexlab{}.
\newblock \showarticletitle{A relational model of authority in groups}.
\newblock In \bibinfo{booktitle}{\emph{Advances in experimental social
  psychology}}. Vol.~\bibinfo{volume}{25}. \bibinfo{publisher}{Elsevier},
  \bibinfo{pages}{115--191}.
\newblock


\bibitem[\protect\citeauthoryear{Wang and Benbasat}{Wang and Benbasat}{2007}]%
        {wang2007recommendation}
\bibfield{author}{\bibinfo{person}{Weiquan Wang} {and} \bibinfo{person}{Izak
  Benbasat}.} \bibinfo{year}{2007}\natexlab{}.
\newblock \showarticletitle{Recommendation agents for electronic commerce:
  Effects of explanation facilities on trusting beliefs}.
\newblock \bibinfo{journal}{\emph{Journal of Management Information Systems}}
  \bibinfo{volume}{23}, \bibinfo{number}{4} (\bibinfo{year}{2007}),
  \bibinfo{pages}{217--246}.
\newblock


\bibitem[\protect\citeauthoryear{Wang, He, Feng, Nie, and Chua}{Wang
  et~al\mbox{.}}{2018}]%
        {wang2018tem}
\bibfield{author}{\bibinfo{person}{Xiang Wang}, \bibinfo{person}{Xiangnan He},
  \bibinfo{person}{Fuli Feng}, \bibinfo{person}{Liqiang Nie}, {and}
  \bibinfo{person}{Tat-Seng Chua}.} \bibinfo{year}{2018}\natexlab{}.
\newblock \showarticletitle{Tem: Tree-enhanced embedding model for explainable
  recommendation}. In \bibinfo{booktitle}{\emph{Proceedings of the 2018 World
  Wide Web Conference}}. \bibinfo{pages}{1543--1552}.
\newblock


\bibitem[\protect\citeauthoryear{Xu, Han, Piao, and Li}{Xu
  et~al\mbox{.}}{2019}]%
        {xu2019think}
\bibfield{author}{\bibinfo{person}{Fengli Xu}, \bibinfo{person}{Zhenyu Han},
  \bibinfo{person}{Jinghua Piao}, {and} \bibinfo{person}{Yong Li}.}
  \bibinfo{year}{2019}\natexlab{}.
\newblock \showarticletitle{" I Think You'll Like It" Modelling the Online
  Purchase Behavior in Social E-commerce}.
\newblock \bibinfo{journal}{\emph{Proceedings of the ACM on Human-Computer
  Interaction}} \bibinfo{volume}{3}, \bibinfo{number}{CSCW}
  (\bibinfo{year}{2019}), \bibinfo{pages}{1--23}.
\newblock


\bibitem[\protect\citeauthoryear{Yaniv and Kleinberger}{Yaniv and
  Kleinberger}{2000}]%
        {yaniv2000advice}
\bibfield{author}{\bibinfo{person}{Ilan Yaniv} {and} \bibinfo{person}{Eli
  Kleinberger}.} \bibinfo{year}{2000}\natexlab{}.
\newblock \showarticletitle{Advice taking in decision making: Egocentric
  discounting and reputation formation}.
\newblock \bibinfo{journal}{\emph{Organizational behavior and human decision
  processes}} \bibinfo{volume}{83}, \bibinfo{number}{2} (\bibinfo{year}{2000}),
  \bibinfo{pages}{260--281}.
\newblock


\bibitem[\protect\citeauthoryear{Zenun~Franco}{Zenun~Franco}{2017}]%
        {zenun2017online}
\bibfield{author}{\bibinfo{person}{Rodrigo Zenun~Franco}.}
  \bibinfo{year}{2017}\natexlab{}.
\newblock \showarticletitle{Online Recommender System for Personalized
  Nutrition Advice}. In \bibinfo{booktitle}{\emph{Proceedings of the Eleventh
  ACM Conference on Recommender Systems}}. \bibinfo{pages}{411--415}.
\newblock


\end{thebibliography}

%









\end{document}